%% file: LehmannForBULL.tex
\documentclass[12pt]{article}
\usepackage{epsfig,pstricks,latexsym}

\newcommand{\be}{\begin{equation}}
\newcommand{\ee}{\end{equation}}
\newcommand{\ba}{\begin{eqnarray}}
\newcommand{\ea}{\end{eqnarray}}

\newcommand{\bd}{\begin{description}}
\newcommand{\ed}{\end{description}}

\newcommand{\nn}{\nonumber}
\newcommand{\system}{{\sc system}}
\newcommand{\ssystem}{semiadditive system}
\renewcommand{\iota}{{\bf 1}}

\def\rellow#1#2{\mathrel{\mathop{\kern 0pt #1}\limits_{#2}}}

\def\S{{\bf S}}
\def\C{\bf C}
\def\U{{\bf U}}
\def\H{{\cal H}}
\def\Alpha{A}

\def\HC{\bf H}
\def\B{\bf B}

\def\X{{\bf X}}
\def\J{{\bf J}}
\def\SigmA{{\mathbf{\Sigma}}}
\newtheorem{ass}{Assumption}

\newtheorem{define}{Definition}
\newtheorem{lemma}{Lemma}
\newtheorem{theorem}{Theorem}
\newtheorem{corollary}{Corollary}
\newtheorem{example}{Example}
\newtheorem{remark}{Remark}

\newtheorem{cat}{Categorical Remark}
\title{Universal Dynamics, a Unified Theory of Complex Systems. Emergence,
 Life and Death\thanks{To be published in Commun. Math. Phys.}}
\author{Gerhard Mack \\
   II. Institut f\"ur Theoretische Physik, Universit\"at Hamburg \\
\date{April 20, 2000}
{\bf Dedicated to the memory of Harry Lehmann}}
\begin{document}
\maketitle
$\quad$\\[-5mm]\noindent 
{\bf abstract}
A universal framework is proposed, where 
all laws are regularities of 
relations between things or agents.
Parts of the world at one or all times are modeled as networks called 
\system s with a minimum of axiomatic properties. 
%
A notion of locality is introduced by declaring some relations direct 
(or links).
Dynamics is composed of basic constituents called 
{\em mechanisms}. They are conditional actions of basic local structural
 transformations (``enzymes''): indirect relations become direct
 (friend of friend becomes friend), links are removed, objects copied. 
 This defines a kind of {\em universal chemistry}.
 I show how to model basic life processes in a self contained fashion
 as a kind of enzymatic computation. The framework
 also accommodates the gauge theories of fundamental physics. 
Emergence creates new functionality by cooperation -
 nonlocal phenomena arise out of local interactions. I explain how this can be understood in a reductionist way by multiscale analysis 
 (e.g. renormalization group).

\section{Introduction}
When we speak about the world, we speak about models of parts of the world
 which are constructed by the human mind. I postulate that they reflect the 
structure of human thinking as formulated in the following

\noindent {\bf preaxiom:}{\em The human mind thinks about relations between things or agents}

Relations will be interpreted as directed binary relations from a source to a 
target. Their constitutive property is that they can be composed - think of 
friend of a friend, brother in law, next nearest neighbor etc.

Traditionally, emphasis in physics has been on objects, like atoms or 
elementary particles. But relations are  equally important. They 
integrate the objects into a network. In adaptive systems, the relations 
change in time in such a way that the connectivity of the whole network may 
change.  Some mistaken views concerning reductionism or emergence result from
neglect of the basic role of relations.
    
One may regard geometry as ancestor of relational theories. It knows a
 relation of parallelism between pairs of tangent vectors which serves to
 define  straight lines and distances. 

The modern theories of fundamental physics are relational theories.
 This is true for the established theories, general relativity and the standard model of elementary particle physics. They are geometric theories 
(gauge theories). It is also true of string theory \cite{strings}
 and of the loop space 
approach to quantum gravity in Ashtekar variables \cite{Ashtekar}.
 Strings, whether open
or closed, can be composed when they touch appropriately, and similarly for 
loops.  

A general plan of {\em relational biology} has been put forward by Rashevsky
\cite{Rashevsky} and Rosen \cite{Rosen} decades ago.

There is a mathematical theory of relations, category theory. Lawvere 
 sought a purely categorical foundation of all mathematics, including set
 theory \cite{Lawvere}.
 The mathematical biologists used category theory from the start. 

However, category theory lacks an essential ingredient of physical theories,
{\em locality}. The fundamental physical theories are local in space time 
in the sense that the basic equations only relate quantities at
 (infinitesimally) close points in space, and at (infinitesimally) close
instances of time. This is the celebrated Nahewirkungsprinzip which was
discovered in the last century. Newton's theory does not obey it, but 
Einsteins general relativity, which supercedes it, does, and so does 
electrodynamics. The processing of chemically bound atoms and molecules in the living cell is mostly performed by biochemical enzymes, and their action is local - they act somewhere at a time. 

 Here I postulate a more general locality principle which 
does not refer to space. Certain relations are singled out as {\em direct
 relations}, called {\em links}, and all others are obtained from them by 
composition. The generalization is desirable for several reasons - 
systems not in space, investigations of properties of space time itself, 
quantum objects in quantum systems
\footnote{They are not considered in this paper}
 whose parts are far apart in space, rapid
communication over long distances (like in the Newtonian limit of general
 relativity) etc. 

The fathers of artificial intelligence did not adopt locality as a default 
option. This lead to such problems as mentioned by Marvin Minsky
\cite{Mentopolis} when he says that a robot needs to be told a lot of facts
about its surroundings, for instance that the wall does not fall down when he
paints the table in the middle of the room. Without a locality principle,
complexity becomes unmanageable \cite{macros}. 

What is assumed is not explained. Therefore,
a fundamental physical theory is the more fundamental the less a priori 
structure is assumed.
 And a theory of complex systems is the more general 
the less a priori structure is assumed. Here I propose a general framework 
which provides a minimum of a priori structure through the definition of a
 \system \ . It is in the spirit of L.v.Bertalanffy
\cite{Bertalanffy}, the pioneer of 
general systems theory, and of Wittgenstein's {\em tractatus} \cite{tractatus}.
 Basically it defines in a precise 
way a notion of structure. Its axioms contain essentially no more than my
 preaxiom and locality. In contrast with automata theory \cite{automata}, 
the framework is supposed to be self contained.
 In principle, there are no data 
in \system s other than  structure, no states of any part of a 
\system \  other than
structure and no information exists other than specification of structure. 
The miracle is how much can be modeled with so little building 
material. I show models of life processes which exemplify this. 
The selfcontainedness of the present framework makes it a natural 
candidate for implementation on a computer. Software has been written and
will be presented elsewhere \cite{Wuerthner}. It offers the convenience that 
one may compose, record and run a program by mouse click, build models
of complex systems in this way and simulate dynamical processes.

With locality, models become much more similar to those which physicists are
used to.  This results in a promising strategy to bring methods of 
theoretical physics to bear on very general complex systems, including
biological and social systems. 

Actually I want to model not only material parts of the world,
 but also space time and immaterial constructs of the human mind like  
proposition logic. Different kinds of \system s are distinguished by 
structural features which generalize what is known in physics as 
{\em constraints} on initial states.
For instance, Gauss' law and gauge group isomorphic to ${\bf R}$ are 
constitutive constraints for an electromagnetic field.
Space embedded in space time is characterized by
constraints which may be summarized by saying that there is 
geometry. And the constitutive property of matter is that it is in
space time. Conservation laws need no be imposed in addition. They follow
 from  requirements of internal consistency. Einsteins equations for space 
time would be inconsistent without covariant conservation of the energy
momentum tensor, and Maxwell's equations of electrodynamics would be 
inconsistent without conservation of electric charge. For reasons of space,
 I cannot expand on these aspects here, cf. ref.\cite{mack:einstein}. 
But note that Darwinian evolution has competition for scarce resources 
as a constitutive feature, and scarcity is  a consequence of
conservation of matter and energy.     

There are no numbers in the framework to begin with, but a numerical 
description can sometimes be obtained by coordinatization. For instance, 
a gauge group is always defined as a structural property of a {\sc System} 
(at one time) and through its coordinatization one obtains a numerical 
description of the  gauge fields  in  lattice gauge theory models of elementary 
particle physics. In fact, the constitutive feature of pure gauge theories
- differential geometry on principal or associated fiber bundles - 
can be recovered from the axioms plus one single extra structural assumption,
{\em forth $\circ$ back = identity}, cf. ref.\cite{mack:faddeev} and section \ref{sec:repres}.

Let us turn to dynamics. I consider local Markovian dynamics in discrete time.
 The dynamics is composed from special local structural transformations. 
They are 
atomic constituents of dynamics and will be called {\em enzymes}. They are
 the  mathematical models not only of the aforementioned biochemical enzymes
, but of any kind
of agent which causes change locally. Besides, there are predicates
which enquire about local structural properties and serve to formulate conditions 
to which the enzymes action is subject. The conditional action of an 
enzyme - i.e. a pair (enzyme, predicate) - is called a {\em mechanism}.
Mechanisms are valuable tools in theoretical immunology 
\cite{schelli}.

It is not a trivial task to construct a well defined deterministic
dynamics from mechanisms
because their actions here and at a neighboring location may not commute. 
Since Petri \cite{Petri} this is known to computer scientists working 
on parallel computing as the {\em concurrency problem}.
I solve it with the help 
of a generalization of the well known device of Jacobi sweeps \cite{MGrid}. 
The resulting theory may be viewed as a {\em universal chemistry} in which
general objects and links substitute for atoms and their chemical bonds. 
\footnote{Actually there is in chemistry another basic relation besides 
chemical bonds, spatial proximity} 
and dynamics can be interpreted as {\em enzymatic computation} \cite{Wuerthner}.
The $\lambda$-calculus \cite{lambda} can be implemented,  
 therefore enzymatic computation can do anything a Turing machine can
\cite{Rathje}.

Enzymes share with matter  the property of being somewhere. In biochemistry
they are tied to material bodies, while in fundamental physics they are 
imagined to be ubiquitous
\footnote{In the canonical approach to classical physics, 
the dynamics it determined by 
the Hamiltonian $H$ according to 
$ \xi \mapsto \xi + \{ H, \xi \}\delta t . $
In a field theory, the Hamiltonian is a sum of local pieces which act locally. 
Call them enzymes. They are composed from basic micro-enzymes -  canonical
variables $q$ and $p$.
We may imagine that these enzymes are everywhere, and they
 stay there. One might want to 
think of them as
dynamical, capable of changing their location or  composition (functional form)
 as a consequence of the
 dynamics.  But that is impossible because  whatever $H$ may be,
the Poisson bracket $\{H , H\}=0$.
  In dissipative systems the situation is different.}.
We  regard the presence of mechanisms as part of the 
specification of the initial state of a \system . 
The universal dynamics says: All mechanisms operate.

A {\sc System} {\em sub specie aeternitatis} - i.e its whole history -   
is again a {\sc system}. It is called a {\em drama}. The dynamics manifests
itself as structural properties of this system. 
 
The dynamics can be stochastic. In this case the drama is a random 
{\sc System}, and its links etc. are random variables. When the dynamics is
sufficiently stochastic, the drama becomes a 
classical equilibrium statistical mechanical system, albeit
 ``in one more dimension'', with possible initial conditions now figuring as
boundary conditions. There may also be external fields which represent 
interactions with an environment. Some emergent phenomena in biological 
systems  can be modeled that way and they then appear as instances of 
familiar phenomena in equilibrium statistical mechanics such as restoration of 
spontaneously broken symmetries. Elsewhere \cite{mack:emergent} I illustrate this on the example of a very
 much simplified model of 
{\em schools of fish swimming in coherent array which abruptly turn
         together with no leader guiding the group.}

Emergence is generally understood as leading to new and often unexpected properties of a whole which are not shared by its isolated parts. 
\footnote{An illuminating discussion of the relevance of concepts in complexity, including emergence, for immunology can be found in I. Cohen's book 
\cite{Cohen}}
We regard a
system as genuinely complex, if its properties are not all shared by subsystems with few objects.
 Emergent phenomena which arise in this way are
nonlocal phenomena. Yet we want to understand them as a consequence of local
 interactions.  Life is an emergent phenomenon. It involves emergent 
functionality. According to Maturana and Varela \cite{autopoiesis}, living
organisms are autopoietic - characterized by being able to make their own
 elements. 
\footnote{ Maturana and Varela thought of biological systems only, but Luhmann
\cite{Luhmann} generalized the notion of {\em Autopoiesis} to social systems. } 
 Typically this involves creation of structure by copying or 
translation from templates,
and preparation of building blocks by digestion,
 i.e. degradation of structure. 
The action of the splitFork-enzyme of section \ref{sec:splitFork} is an 
example of emergent functionality. The whole enzyme can copy arbitrary 
\system  s, but the individual micro-enzymes from which it can be composed
 cannot. And when one of them is missing or carries the wrong predicate,
the copy-functionality is also lost. 

Some emergent phenomena like wave propagation can be understood by 
exploitation of symmetry and linearity or other special methods. Otherwise,
a multiscale analysis is called for. Although a genuinely complex
system $\S$ cannot be understood as a whole by looking at it locally, 
a complexity reduction is often possible by local considerations. This was
the central idea of Wilson's renormalization group \cite{RG,RG2}.
 One constructs 
an effective theory, i.e. a 
description in terms of a new system $\S^1$ with new objects
 which represent subsystems, but retain only as much information on
their internal structure as is relevant for their cooperation. 
Links between them are also constructed. Enzymes may be attached 
which represent functionality of compound enzymes at the smaller scale.     
The resulting system is still complex, but may have much fewer degrees
of freedom. Then the procedure of complexity reduction may be iterated, 
leading to a multilevel description. 

Mathematically, the chief insight is the relation between  block spin
constructions in a renormalization group (RG) setup, and collections of
dual colimits and factorizing cones  in categories. Experience with the 
rigorous renormalization group approach to gauge theories 
\cite{Balaban} is valuable 
because general \system s share many features of lattice gauge theories,
cf. section \ref{sec:repres}.
In what way they are essentially more general is best seen in
the section \ref{sec:logic} on logic.
 Monte Carlo RG-studies of gauge theories have also been performed, cf. e.g. 
\cite{gupta:RG}, and the Monte Carlo RG-method is still being improved
\cite{brandt:RG}. 
\section{Structure}
This section and the next introduce the basic mathematical framework.
Parts of the world at one time as well as their whole histories are modeled
as \system 's with certain axiomatic properties. A more descriptive name 
would be ``local category''.

According to L. van Bertalanffy, \cite{Bertalanffy},
{\em a system is a set of units with relationships between them.}
I precisize. 

\begin{define}\label{def:system}{\em ({\sc System})}
A \system \ is a 
model
 of a part of the world as a network of
{\em objects} $X,Y,...$ (which represent things or agents) with {\em arrows} 
$f,g,... $ which represent directed relations between them.\\ One writes 
$f:X\mapsto Y$ for a relation from a {\em source} (domain) $X$ to a
 {\em target} (codomain)
$Y$.

  The arrows are characterized by axiomatic properties as follows:
\begin{enumerate}
\item 
{\em composition}. Arrows can be composed.
If $f:X\mapsto Y$ and $g:Y\mapsto Z$ are arrows,
then the arrow $$g\circ f: X \mapsto Z$$ is defined.
The composition is associative, i.e. $(h\circ g)\circ f= h\circ (g\circ f)$. 
\item
{\em adjoint}.
To every arrow $f: X\mapsto Y $ there is a unique arrow
$f^{\ast }:Y\mapsto X $ in the opposite direction, called the 
adjoint of $f$. $f^{\ast\ast} = f$ and
$(g\circ f)^{\ast}=f^{\ast}\circ g^{\ast}$. 
\item
{\em identity}. To every object $X$ there is a unique arrow 
$\iota_X: X\mapsto X $ which represents the identity of a thing or agent
 with itself. 
$$\iota_X = \iota_X^{\ast}, \qquad \mbox{and } \quad \iota_Y\circ f =f= f\circ \iota_X$$
for every arrow $f:X\mapsto Y$.  
\item
{\em locality}: 
Some of the arrows  are declared      
{\em direct} (or fundamental); they are called {\em links}.
 All arrows $f$ can be 
made from links by composition and adjunction,
 $f=b_n\circ ... \circ b_1, (n\geq 0)$ where $b_i$ are links or
adjoints of links; the empty product $(n=0)$ represents the identity. 

\item
{\em composites:} The objects $X$ are either {\em atomic} or {\sc System}s.
In the latter case, $X$ is said to  have internal structure,
 and the objects  of the \system \ $X$ are called its {\em constituents}. 

\item{\em non-selfinclusion:}
A \system \ cannot be its own object or constituent of an object etc. 
Ultimately, constituents of ... of constituents are atomic. 

\end{enumerate}
The links and objects of a system are called its {\em elements}. 
\\
A system is called {\em connected} if there are arrows to all other objects
from some (and therefore all) objects. 
\\
A system is called {\em unfrustrated} if there is at most one arrow 
from $X$ to $Y$ for any objects $X$, $Y$.
\end{define}

Axioms 1 and 3 are those of a category. Ignoring specification of links and
the $*$-operation, a \system \ $\S$ becomes a category $Cat(\S)$. 
 
There is a long standing controvercy in philosophy concerning identity,
see e.g. Wittgenstein \cite{tractatus}, Satz 5.303 or Quine \cite{Quine}.
 In \system s theory, the identity arrows $\iota_X$ are as important as 
the number 0 in arithmetics. Later we shall have occasion to introduce also
special arrows between two objects which are identical in the sense of 
indistinguishable (i.e. copies). By abuse of language they will also be called
{\em identity arrows}.

 The idea of an adjoint (axiom 2) is that 
a relation in the opposite direction should be specified in some way by
any link. There can be different ways in which links can be adjoint. For 
instance, if objects $X$, $Y$ are {\sc System}s, hence categories, and
 $f^\ast :Y\mapsto X $ is left adjoint functor of a functor $f$ (s. later)
 then $f^{\ast \ast}=f $ is right adjoint functor of $f^\ast$.  

 Axiom 4 introduces locality as explained in the introduction.
 
 Axiom 5 makes the whole scheme self contained. And according to Jacob 
\cite{Jacob} {\em Tout objet que consid\`ere la biologie represente un syst\`eme de syst\`emes.}

The totality of statements about {\sc system}s which are meaningful 
as a consequence of the axioms will be called the
{\em language of thought}.

\begin{ass}
\label{stAss} Unless otherwise stated, it is assumed that constituents, 
constituents of constituents etc. of objects of $\S$ are not objects of $\S$.
\end{ass}

It must be emphasized that
atomicity of an object is not a property of something in the world, but  
of a particular model which describes some of its aspects on some scale. 

Objects with internal structure are {\em black boxes}.
Later on we shall ``dissolve'' such objects, making their interior structure 
visible by putting links from some of their constituents, and thereafter they 
may be treated as atomic although they still stand for the ``same''
(composite) object. 
\begin{define}{\em (Types of links)}
A link $b :X\mapsto Y$ is 
said to be invertible if there exists an arrow, denoted $b^{-1}$
such that $b\circ b^{-1} = \iota_Y$, and 
$b^{-1} \circ b = \iota_X$.

It is said to be {\em unitary} if $b^\ast = b^{-1}$.

Links whose adjoints are links will be called {\em bidirectional}
 for short.
\end{define}

Sequences $b_1,...,b_n$ of links or adjoints of links which can be composed 
are said to make {\em a path of length} $n\geq 0$.
 Composability requires that 
the target of $b_i $ is the source of $b_{i+1}$. The length of the shortest
connecting path can be used to measure distance between objects. 
   
\begin{define} {\em (Subsystems)}
A {\em subsystem} $\S_1$  of a \system \  $\S$ is generated by a set of 
objects in $\S$ 
and a set of links in $\S$  between these objects.
Its arrows are all arrows in $\S$ that can be composed from these links and
their adjoints. 

The {\em boundary} of $\S_1$ consists of the links in $\S$ with target in 
$\S_1$ which are not links or adjoints of links in $\S_1$. 

The {\em environment} of $\S_1$ is the system generated by the objects of 
$\S$ not in $\S_1$ and the links between them.

For $n\geq 1 $, the $n$-neighborhood of an object $X$ is the subsystem 
which contains all  objects connected to $X$ by a path of length $\leq n$, 
and is generated by all links between them. Identity links in the path 
are counted as contributing 0 to its length. 

A {\sc System} is called \emph{locally unfrustrated} if all its
$1$-neighborhoods are unfrustrated subsystems.   
\end{define}
Note that it is {\em not}  required that adjoints of links in $\S_1$ which are
links in $\S$ are also links in $\S_1$.  

Following Luhmann \cite{Luhmann}, one may also want to consider 
the {\em internal environment} ${\bf I}_1$ of $\S_1$. 
It is the system whose objects are
the constituents of non-atomic objects $X$ in $\S_1$ and
 whose links are the links between 
them.  Arrows are equivalence classes of paths, equivalences
being determined by the equivalences in the \system s $X$. Given our standing
 assumption \ref{stAss}, ${\bf I}_1$ is {\em not}
a subsystem of $\S$.

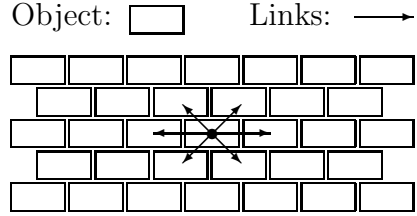
\begin{figure}
\begin{center}
\begin{picture}(180,100)(0,0)
%
%
%
\put(10,80){\mbox{Object:}}
\put(55,77){\frame{\makebox(20,10){}}}
\put(100,80){\mbox{Links:}}
\put(140,83){\vector(1,0){22}}
%
%
\multiput(10,10)(22,0){7}{\frame{\makebox(20,10){}}}
\multiput(10,34)(22,0){7}{\frame{\makebox(20,10){}}}
\multiput(10,58)(22,0){7}{\frame{\makebox(20,10){}}}
\multiput(20,22)(22,0){6}{\frame{\makebox(20,10){}}}
\multiput(20,46)(22,0){6}{\frame{\makebox(20,10){}}}
%
%
\put(86,39){\circle*{4}}
\put(86,39){\vector(1,0){22}}
\put(86,39){\vector(-1,0){22}}
\put(86,39){\vector(1,1){11}}
\put(86,39){\vector(1,-1){11}}
\put(86,39){\vector(-1,1){11}}
\put(86,39){\vector(-1,-1){11}}
\end{picture} \end{center}
\caption{Brick wall. The links $b$ are translations of a brick to a 
nearest neighbor's position. The arrows are equivalence classes of paths 
subject to the equivalences $b^\ast = b^{-1}$ and $b_1\circ b_2\circ b_3 = 
\iota_X$ for any triangular path from  $X$ to $X$. The system is locally
unfrustrated, but would become (globally) frustrated if the wall were
 closed to a round tower. }\label{fig:brickWall}
\end{figure}

The brick wall shown in figure \ref{fig:brickWall} is an example of a locally 
unfrustrated system, and so is any triangulated 2-manifold,
 with the 1-simplices as links. 

\subsection{Structure preserving maps}
As always in physics, we shall not distinguish between isomorphic systems. 
To make this precise we need to consider structure preserving maps 
called {\em local functors}. 
\footnote{This makes precise what Wittgenstein leaves undefined 
in his isomorphism theory 
 in {\em tractatus } 2.15 when he postulates that the images of two 
 objects are
related ``in the same way'' as the objects.}

In category theory, a {\em functor } $F: \C \mapsto \C^\prime$ is a map
of the objects of a category $\C $ to objects of a category $\C^\prime $ and
 of arrows $f$ of $\C $ to arrows of $\C^\prime $ such that source and target
of $F(f)$ are the images of source  and target of $f$, and
\ba
F(f\circ g) &=& F(f) \circ F(g), \label{Functor:1}  \\
F(\iota_X) &=&
\iota_{F(X)} \ . \label{Functor:2}
\ea
In contravariant functors, eq.(\ref{Functor:1}) is replaced by
\be
F(f\circ g) = F(g) \circ F(f), \label{Functor:3}  \\
\ee
\begin{define}{\em(local functor)}
A local functor $F: \S \mapsto \S^\prime $ is a map of a \system  \ $\S$ 
into another \system \ $\S^\prime $ which obeys the above requirements on 
a functor of categories, maps links into links, and obeys
\be
F(f^\ast) = F(f)^\ast
\label{Functor:4}
\ee
In a contravariant local functor, eq.(\ref{Functor:3}) is substituted for
eq.(\ref{Functor:1}). 

An {\em isomorphism} of \system s is a local functor whose inverse exists 
as a local functor. An {\em anti-isomorphism} 
of \system s is a contravariant   local functor whose inverse exists 
as a contravariant local functor.
\end{define}
No local functor between two \system s need  exist unless the identity arrows 
of the image are links. (Typically they are not.)

Anti-isomorphisms relate complementary shapes.
They are important in cognition.  The surface
(boundary) of a lock and a key have opposite orientation; this will be 
reflected in an anti-isomorphism. This is important for life.
In biochemistry, the specificity of enzymes for particular substrates 
is due to a lock-key match of parts of their surfaces. 
And receptors on cell walls function according to the same lock-key principle
\cite{Alberts}.

It is important to note that the internal structure of black boxes 
(nonatomic objects) is declared irrelevant by not distinguishing isomorphic
\system s, and so is the distinction between atomic and nonatomic objects. The
only usage of the internal structure is in constructing 
links of the \system \ and their composition $\circ$. One does not look into 
black boxes anymore once they are in place. 

If isomorphism of corresponding nonatomic objects is demanded, we speak of
a {\em strong isomorphism}. 

\begin{cat}
Local functors define the category of \system s whose objects are \system s
and whose arrows are local functors. 
\end{cat} 
Among the isomorphisms $F: \S \mapsto \S^\prime$,
there is a particularly important class called {\em gauge transformations}.

 They are ``inner'' in the sense that they are generated by 
arrows of the system. 
Since gauge transformatons are isomorphisms, 
statements about a \system \ in the language of
thought are necessarily
gauge invariant. It follows that observables must be gauge invariant. 

More precisely, one may define objective observables to be  boolean functions 
on \system s, well defined for all \system s, while subjective 
observables require for their definition specification of a distinguished 
object $X$ (the subject), and possibly some links $b$ with target $X$
(e.g. the direction in which a speaker points). 
Objective observables must be gauge invariant, while for subjective observables
this needs only be true for gauge transformations which are trivial at $X$
and don't change its specified links $b$. 

Given $\S$ and an object $X$ of $\S$, 
the group $G_X$ of local gauge transformations consists of all invertible
arrows $g : X\mapsto X $. (It is also called the holonomy group).
Its identity element is $\iota_X$.  

Gauge transformations take $\S$ into a system $\S^\prime $ with
$Cat(\S)$$=Cat(\S^\prime)$. 

\begin{define}{\em (gauge transformations)}
 A gauge transformation $F$ is defined by selecting 
an element $g_X \in G_X$ for every object $X$, and mapping
arrows $f:X\mapsto Y$ into 
\be  F(f) = g_Y^{-1}\circ f \circ g_X \label{gaugeTransf:def} . \ee
The links of $\S^\prime$ are the images of links of $\S$, and the 
 *-operation ${}^+$ in $\S^\prime$ is defined in terms of
 the *-operation in $\S$ by
 \be f^+ =  \sigma_X^{-1}\circ f^\ast \circ \sigma_Y  \label{gaugeTransf:*op}\ee
with $\sigma_X = g_X^\ast \circ g_X$.
A gauge transformation is called {\em unitary} if
 $g_X^\ast=g_X^{-1}$ for all $X$.
\end{define} 
One verifies that the conditions for an isomorphism are satisfied. 
The composition law is preserved. The new star operation 
satisfies $f^{++}=f$ and $\iota_X^+=\iota_X$ as it should, and 
$F(f^\ast)=F(f)^+$.

\begin{theorem}{\em (Gauge group)} In connected systems $\S$ where all links are
 unitary, all gauge transformations are unitary, and all groups 
$G_X$ of local gauge transformations are isomorphic. Their isomorphism class 
$G$ is called the {\em gauge group}. 
\end{theorem}
{\em Proof:} With all links, all arrows are unitary. Therefore 
all $g_X:X\mapsto X$ are unitary. Given $X,Y$, there exists an arrow 
$f:X\mapsto Y$ because $\S$ is connected. If $g_X\in G_X$ then 
$g_Y = f\circ g_X \circ f^\ast \in G_Y$, and conversely.  Therefore
$G_X$ and $G_Y$ are isomorphic. q.e.d.

Gauge transformations in electrodynamics are the standard example.
 Electrodynamics will be considered in section \ref{sec:drama} later on.
 Here is another example  

 \begin{example} {\em (Fundamental group)}
  Consider the system made from a 
triangulated manifold as follows. The objects are the 0-simplices 
and the links are the 1-simplices; adjunction is change of orientation,
adjoints of links are links.
 The arrows are 
 equivalence classes of paths $b_0,...,b_n$, defined by the following 
equivalence  relation:

- all links are unitary, 
 
- $b_0\circ b_1\circ b_2 = \iota_X $ if $b_0,b_1,b_2$ is a closed path from 
$X$ to $X$ (a triangle).
\\
The gauge group is the fundamental group of the manifold. 
\end{example}

\begin{theorem}
{\em (Reconstruction of objects)}
\label{theo:objectRec}
A \system \ is determined up to isomorphism if 

- the arrows are enumerated or
given as a set, 

- it is specified which arrows can be composed and which arrow
is the result

- the links among the arrows are specified.

- the adjoints of arrows are specified.

Conversely, any such data determine a \system \ if all the arrows can be 
obtained by composing links and their adjoints, and if the *-operation
satisfies the consistency conditions imposed by the axioms.  
\end{theorem}
{\sc Proof} The corresponding result in category theory is standard \cite{CWM}.
One reconstructs the objects as equivalence classes $X$ of links, two 
links $b_1$ and $b_2$ being equivalent if there exists an arrow $f$ 
such that $b_1\circ f$ and $b_2\circ f $ are both defined, $X$ is then the 
common target of all links equivalent with $b_1$. $f$ with target $X$ 
is equal to the identity $\iota_X$  if 
$g\circ f = g $ whenever it is defined. 

\begin{cat}
Gauge transformations are invertible functors 
 which preserve objects and admit a natural transformation to the identity. 
\end{cat}

\subsection{Tautological character of the axioms}
\label{sec:tauto}
I wish to convince the reader that there is no more in the axioms than 
what is intended by the preaxiom, and locality. The composition law
and adjunction need to be discussed. 

The axiomatic properties assure that a {\sc system} is a directed 
(pseudomulti) graph
if  the number of objects is 
finite. The edges of the graph are the links
and its vertices are the objects. 
The graph may have edges which are loops (pseudographs), and it can have
 multiple edges between the same 
vertices (multigraph)
\cite{graph:gould}.
Deviating from standard nomenclature, I will also speak of a graph
when the number of vertices is countable. 

Let us first turn to adjoints. One may consider the existence of the 
fundamental relation $b:X\mapsto Y$ as a directed relation from $Y$ to $X$. 
This amounts to introducing  formal adjoint links in the graph in the opposite
direction. 

Given a graph $\Gamma $, 
which represents basic relations between 
things or agents, we have no composition rule. But 
one can make 
from  $\Gamma $ a category $\S_\Gamma $ in a universal way. This yields a 
\system . The arrows from $X$ to $Y$ are the paths from $X$ to $Y$ made
from links and their formal adjoints.  
All the \system s with given directed graph $\Gamma $ are obtained 
from $\S_\Gamma $ by 
passing to equivalence classes of paths. So, {\em  the freedom of choosing the composition law
 merely introduces the option  of waiving distinctions between relations. }

Similarly, there is a universal way of making a category
$\U_{\Gamma}$  with unitary links
from any given directed graph. It is obtained from $\S_\Gamma$ by 
imposing the relations $ b\circ b^\ast = \iota $ and $b^\ast \circ b = \iota$
i.e. considering only non-backtracking paths. 
Every \system \ with unitary links is obtained from $\U_\Gamma$ by passing to 
equivalence classes of non-backtracking paths. 

\begin{cat}
The passage to equivalence classes of arrows defines a unique local functor. 
Therefore we have   
\end{cat}
\begin{theorem}
The forgetful functor $F$ from the category of finite \system s $\S $
$[$resp. finite \system s with unitary links $]$
to the category of 
directed (pseudomulti-)graphs $\Gamma $ has a left adjoint functor 
$F^\ast: \Gamma \mapsto \S_\Gamma \ [\mbox{resp.} \Gamma \mapsto \U_\Gamma ]$. 
\end{theorem}

\section{Universal dynamics}
Dynamics shall be compose from local structural transformations. They
are special graph transformations \cite{graphTransf}
 obeying some strict locality
requirements.
They are universal in the sense that their action is defined for arbitrary
systems
 the same will therefore be true for dynamics composed from
 them\footnote{Universal dynamics is intended to be universal also in the sense of universal constructions in category theory. But the appropriate
theorems have not been proven yet.}.
 
 We shall distinguish four kinds of such transformations

{\em motion

growth

death

cognition.}

They are reversible except for death. I discuss them one by one. 

{\em Motion} promotes indirect relations to direct relations.
Either arrows composed of two (or more) links are promoted to the status
of a link (e.g. friend of a friend becomes friend), 
or a non fundamental adjoint becomes a link. The opposite,
 demotion of an adjoint 
link to non-link is also subsumed under motion.  
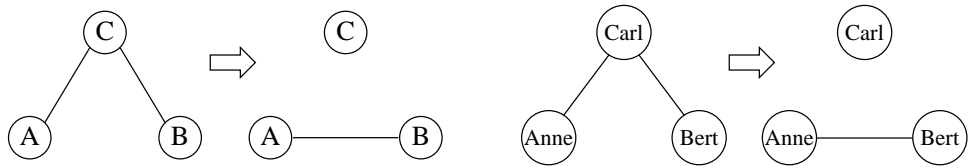
\begin{figure}[b!]
  \begin{center}
    \leavevmode
     \input{catalyse.pstex_t}
  \end{center}
  \caption{\small Catalysis in chemistry and elsewhere. A catalyst {\tt C} binds molecules {\tt A} and {\tt B}. First a substrate-enzyme complex is built, where {\tt A} and {\tt B} bind to {\tt C}. Next the composite arrow from
{\tt A} to {\tt B} becomes fundamental.}
\label{fig:catalysis}
\end{figure}

Equations 
of motion in physics like the Maxwell equations (s. below) determine motion in
this sense. Catalysis of bonds in chemistry (and elsewhere, 
figure \ref{fig:catalysis}) is motion in this sense (supplemented with removal 
of some link(s)), and so is motion in space. Let material body $A$ be ``at'' 
space point $C$ and $C$ be ``neighbor$\_$of'' $B$. If the  
relation ``at${}^\prime =$neighbor$\_$of $\circ$ at'' becomes fundamental 
instead of ``at'', it means that body $A$ has moved from $C$ to $B$.       

 The category  $Cat (\SigmA_t)$  does not change at all
in this kind of time evolution. Therefore we have
\begin{theorem}
 Any quantity $Q$ which is 
determined by the category $Cat(\SigmA_t)$ is a constant of \emph{motion}.
\end{theorem}

{\em Growth} copies objects.

{\em Death} removes links (together with their adjoints), or removes objects 
together with all links incident on them.

{\em Cognition} creates links between objects with matching internal
 structure. The match is supposed to be established by enzymatic computation,
ultimately with enzymes of the other kinds. 
 Identity links (cf. after definition \ref{def:system})
between atomic copies of objects are also admitted 
and serve as prototypical examples of cognitive links. 
The creation of links by cognition (cognitive links for short) is fundamentally
different from creation of links from existing links
by composition or adjunction because it creates {\em new} arrows. 
This will be important later on when we adapt Baas' distinction between
deductive and observational emergence \cite{Baas:1}. 
In principle,  operations of this kind would also be
mathematically well defined if they are nonlocal, but we don't want to admit 
that. If only objects
within one 1-neighborhood can be linked, they must be connected before by 
a path of length at most 2. But the new arrow is not the arrow which
this path defines.
For instance, in chemistry spatial proximity links are a prerequisite 
for forming chemical bonds. Receptors on membranes of living cells operate
in a cognitive way, using lock-key type matching.   

Before proceeding to formal developments, I will discuss some examples. 

Copy processes are most important for life. 
Autopoietic systems make their own elements. Ultimately they make them from
constituents or constituents of constituents etc. which are 
conserved material entities, not created. So the ``making'' is
supply of structure, typically from templates. DNA is copied, 
it is also copied 
into RNA. RNA is copied and is also translated into sequences of amino acids,
i.e. proteins.  

The material constituents must also be supplied. 
Plants make organic material from inorganic substances and light by 
enzymatic action, but animals need to get their building blocks from organic
 materials  breaking down its preexisting structure.
Call this digestion. 

 I will demonstrate how 
copy processes and digestion can be modeled using no more than what is 
provided by the axioms, definition \ref{def:system},
 and also how relaxation sweeps
through extended \system 's can be modeled.  
\subsection{Examples}
\label{sec:examples}
\subsubsection{Copying. The asymmetrical replication fork}
\label{sec:splitFork}
There is an example of a local dynamics that 
 can be used to produce within a finite time 
two copies of any finite \system \ whose links are all bidirectional.
It is a mathematical abstraction and generalization of the asymmetrical 
replication fork mechanism which copies DNA in the living cell, see
standard text books \cite{Alberts,genes}.
 During the copy process, links without 
fundamental adjoints appear. 

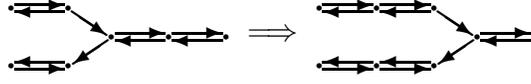
\begin{figure}
\label{fig:splitFork}
\begin{center}
\setlength{\unitlength}{0.001875in}%
\begin{picture}(1474,180)(123,460)
\thicklines
\put(1310,470){\circle*{14}}
\put(1310,630){\circle*{14}}
\put(1160,640){\vector( 1, 0){140}}
\put(1000,460){\vector( 1, 0){140}}
\put(1150,470){\circle*{14}}
\put(1150,630){\circle*{14}}
\put(1000,640){\vector( 1, 0){140}}
\put(1140,620){\vector(-1, 0){140}}
\put(1140,480){\vector(-1, 0){140}}
\put(1300,620){\vector(-1, 0){140}}
\put(1580,540){\vector(-1, 0){140}}
\put(1440,560){\vector( 1, 0){140}}
\put(860,540){\makebox(0,0)[b]{\smash{$\Longrightarrow$}}}
\put(990,470){\circle*{14}}
\put(1300,480){\vector(-1, 0){140}}
\put(1160,460){\vector( 1, 0){140}}
\put(990,630){\circle*{14}}
\put(1590,550){\circle*{14}}
\put(1420,540){\vector(-3,-2){ 96.923}}
\put(130,470){\circle*{14}}
\put(140,640){\vector( 1, 0){140}}
\put(280,620){\vector(-1, 0){140}}
\put(300,620){\vector(3, -2){ 96.923}}
\put(410,550){\circle*{14}}
\put(130,630){\circle*{14}}
\put(290,470){\circle*{14}}
\put(290,630){\circle*{14}}
\put(570,550){\circle*{14}}
\put(730,550){\circle*{14}}
\put(400,540){\vector(-3,-2){ 96.923}}
\put(560,540){\vector(-1, 0){140}}
\put(580,560){\vector( 1, 0){140}}
\put(1430,550){\circle*{14}}
\put(1320,620){\vector(3, -2){ 96.923}}
\put(720,540){\vector(-1, 0){140}}
\put(280,480){\vector(-1, 0){140}}
\put(140,460){\vector( 1, 0){140}}
\put(420,560){\vector( 1, 0){140}}
\end{picture}
\end{center}
 \caption{Action of the splitFork-enzyme $s_X$ at $X$, for chains like DNA. 
The same  mechanisms can operate on general \system s}
 \end{figure}

 A {\em fork } at $X$ shall be a pair
of links without fundamental adjoints with source and target $X$,
respectively. The splitFork-action $s_X$ shall be 
a local structural transformation. Figure \ref{fig:splitFork} shows its action on  chains (of pairs of directed links) like the DNA double helix. 
The generalization is as follows. A link is called ``bidirectional'' if its 
adjoint is a link, and ``unidirectional'' otherwise. 
\begin{enumerate}
\item A copy $X^\prime $ of $X$ is made.
\item The links incident on $X$ other than loops are distributed among $X$ and its copy as follows:
\begin{itemize}
\item[-] bidirectional links with target $X$ get $X^\prime$ as their target
\item[-] unidirectional links with target $X$ retain $X$ as their target
\item[-] bidirectional links with source $X$ retain $X$ as their source
\item[-] unidirectional links with source $X$ get $X^\prime $ as their source
\end{itemize}
The loops $X \mapsto X$ remain in place
 and get a copy $X^\prime \mapsto X^\prime$.
\item The adjoints of formerly unidirectional links are promoted to the status
of links. 
\end{enumerate}

\begin{theorem}{\em (Universal copy constructor)} \label{theo:cpy}
Let $\S_0$ be obtained from a finite connected system $\S$ whose links are 
all bidirectional
by action of $s_{X_0}$ at some $X_0\in \S$. For $t>0$, let $\S_t $
be obtained from $\S_{t-1}$ by action of $s_X$ for all objects $X$ which have
forks. 

$\S_t$ is well defined for $t\geq 0$. For 
sufficiently large $t$, it is independent of $t$ and
consists of two disconnected \system s, both isomorphic to $\S$.
\end{theorem}
``Once replication has started, it continues until the entire \system \ has 
been duplicated''. Upon substituting ``genome'' for ``\system `` this becomes 
a quote from a genetics text book \cite{genes}.  

\begin{remark}
 Copying may be initiated at several sites $X_0$,... $X_n$ which are not 
connected by links. 

The action of the splitFork-enzyme is quite robust against 
errors due to computer failures which mimic local mutations.
But  a third copy is made of part of the \system \ when a fundamental adjoint
 gets lost (or added) ``at the wrong moment''
\footnote{In man, errors in copying the genome may result in Down's syndrome,
the presence of three copies of chromosome no. 21 instead of the usual two}.
\end{remark}

The theorem  was first demonstrated in \cite{mack:kyoto}. The fact that 
$s_X$ is well defined requires a comment - what does it means that 
``a copy $X^\prime$ of object $X$ is made''?
 Theorem \ref{theo:objectRec} can be invoked to describe $s_X\S$ up to 
isomorphism. The isomorphism class does not 
retain the information about the internal structure of non-atomic objects.
 But this information can be retained by the copies, if desired. To do so,
one uses the universal copy constructor to copy the objects which are themselves \system s, to copy their non-atomic constituents, and so on. 
 Because of the axiom of non-self-inclusion this does not lead to an infinite
 recursion. In conclusion,
if $s_X$ is defined in this way, the stronger version of theorem \ref{theo:cpy}
holds true where the phrase ``both isomorphic to $\S$'' is replaced by 
``both strongly isomorphic to $\S$''.

$s_X$ is an example of a local structural transformation (``enzyme'').  
On \system s as occur as $\S_t$ in theorem
\ref{theo:cpy}, actions $s_X$, $s_Y$ commute. But this is a lucky circumstance.

The production of a copy is an ``emergent phenomenon'',
a nonlocal phenomenon which arises from local interactions. New functionality 
-copying- emerges.

\begin{cat}
$s_X$ can be decomposed into several ``micro-enzymes'' which act in sequence.
If all links are unitary, each of them specifies a functor of categories, 
albeit the trivial one, but the first of these functors is not surjective 
because an extra object is created. 
\end{cat}
 Such situations are admitted in the 
 work of Ehresmann and Vanbremeersch \cite{EV} to which we turn later.
 It is not true  that 
the functor {\em is} the dynamics in their work.

The decomposition is shown in ref.\cite{Wuerthner}. 
The first micro-enzymes is an elementary copy process
which creates a duplicate of $X$ and links it by an identity link to the 
original. The other micro-enzymes compose links and remove the identity link again. 

 When the copied object
$X$ is an indestructible material constituent,
 its equal may be imagined to be recognized in the environment 
(by a cognitive process) 
and absorbed by linking it to $X$ by an identity arrow. .

\subsubsection{Digestion}
\label{sec:digestion}
Consider a finite connected system $\S_0 $ with bidirectional links and with a 
distinguished object $X$. I describe a local structural transformation 
$d_X$ which continues to act on a 1-neighborhood of $X$. Starting with 
an arbitrary $\S_0$, the action becomes trivial after some time and there 
results a system $\S $ with the same objects 
$X$, $Y_1,...,Y_n$ as $\S$ but whose structure is completely 
degraded in the sense that its only links are one link from $Y_i$ to $X$ 
for each $i$, and their adjoints. (These links could be removed, but then 
``the food $\{ Y_i \}$ is lost'' since there are  no relation to it 
anymore.) 

$d_X$ consists of consecutive steps.

1. (Death) The far side of all triangles of 3 links with tip $X$ is removed, 
together with its  adjoint. 

2. (Motion)
 If $b$ is a link from $Y\neq X$ to $X$ and $b^\prime $ is a link from $
Y^\prime $ to $Y$ then $b\circ b^\prime $ becomes a link and $b^\prime $
ceases to exist as a link. 

3. Fundamental loops $X\mapsto X$ are removed.

\begin{figure}
\label{fig:digest}
\begin{center}
\setlength{\unitlength}{0.7mm}
\begin{picture}(40,50)(40,0) 
\put(0,0){\begin{picture}(0,0)
\put(20,0){\circle*{2}}
\put(0,20){\circle*{2}}
\put(20,20){\circle*{2}}
\put(0,40){\circle*{2}}
\put(20,40){\circle*{2}}

\put(20,0){\line(-1,1){20}}
\put(0,20){\line(1,0){20}}
\put(20,20){\line(-1,1){20}} 
\put(0,20){\line(0,1){20}}
\put(20,0){\line(0,1){40}}
\put(0,40){\line(1,0){20}}
\put(10,0){$X$}
\put(28,0){$\Rightarrow$}
\end{picture}}

\put(30,0){\begin{picture}(0,0)
\put(20,0){\circle*{2}}
\put(0,20){\circle*{2}}
\put(20,20){\circle*{2}}
\put(0,40){\circle*{2}}
\put(20,40){\circle*{2}}

\put(20,0){\line(-1,1){20}}
\put(20,20){\line(-1,1){20}} 
\put(0,20){\line(0,1){20}}
\put(20,0){\line(0,1){40}}
\put(0,40){\line(1,0){20}}
\put(10,0){$X$}
\put(28,0){$\Rightarrow$}
\end{picture}}

\put(60,0){\begin{picture}(0,0)
\put(20,0){\circle*{2}}
\put(0,20){\circle*{2}}
\put(18.8,20){\circle*{2}}
\put(0,40){\circle*{2}}
\put(20,40){\circle*{2}}

\put(20,0){\line(-1,1){20}}
\put(20,0){\line(0,1){40}}
\put(19,0){\line(0,1){20}}
\put(20,0){\line(-1,2){20}}
\put(0,40){\line(1,0){20}}
\put(10,0){$X$}
\put(28,0){$\Rightarrow$}
\end{picture}}

\put(90,0){\begin{picture}(0,0)
\put(20,0){\circle*{2}}
\put(0,20){\circle*{2}}
\put(18.8,20){\circle*{2}}
\put(0,40){\circle*{2}}
\put(20,40){\circle*{2}}

\put(20,0){\line(-1,1){20}}
\put(20,0){\line(0,1){40}}
\put(19,0){\line(0,1){20}}
\put(20,0){\line(-1,2){20}}
\put(10,0){$X$}
\end{picture}}

\end{picture} 
\end{center}
\caption{Digestion enzyme attacks at $X$}
\end{figure}
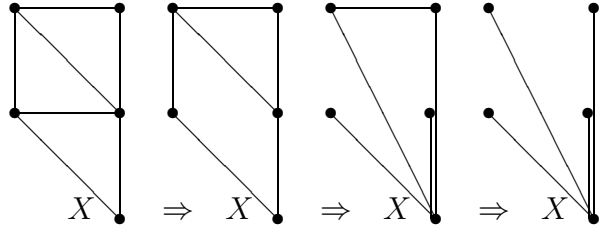

Actually the 3rd step can be omitted when step 2 operates also for 
$Y=X$. 
Figure \ref{fig:digest} illustrates the procedure.  

\begin{cat}
$d_X$ specifies a functor of categories, albeit the trivial one, if
the links are invertible, e.g. unitary. 
\end{cat}
%

%
\subsubsection{Sweeps through a \system}
The splitFork copy procedure of section \ref{sec:splitFork} relies on the 
propagation of a shock wave. The shock wave is the boundary between the part
 of the system which has already been copied, and the rest. It is made of 
links without fundamental adjoints. 
\footnote{Boundaries of subsystems which are 
distinguished in this or other ways serve as mathematical models of membranes 
in cell biology \cite{Alberts}}
The objects $X$ at the outside of the boundary are copied, and then the 
boundary passes past them. Instead of copying $X$, one may act on its
neighborhood  with some (other)  enzyme. In this way a sweep through the
 \system\ is generated which invokes an updating of the neighborhoods of all 
objects in the \system .
\subsection{Systems {\em sub specie aeternitatis}}
\label{sec:drama}
There are two different ways of looking at dynamics. I will now examine
 the possibility of looking at a system {\em sub specie aeternitatis}, i.e. 
its whole history. This will also be viewed as a \system. Following 
Sorin Solomon's suggestion, I call it a {\em drama}. 
  
\begin{define} {\em (Drama)} A drama is a system $\S$ which is composed from
subsystems $\S_t$ labeled by $t=0, \pm 1, \pm 2, ... $ and links $e$ 
in one or both directions between objects 
$Y \in \S_{t+1}$ and $X \in \S_t$. If there is such a link, 
$Y$ is said to be descendent of $X$, 
and $X$ is an ancestor of $Y$. It is required that 
every $Y\in \S_{t+1}$  is descendent of at least one object in $\S_{t}$. 
If there are several such objects, they must be connected by identity links. 
\end{define}
We will impose the additional condition that the time links are all unitary, 
and  that any two paths between the same target $X$ and source $Y$ which are
 made exclusively of time links define the same arrow. It is a subtle question 
whether one might want to generalize this. If the condition is satisfied,
the time links can be ``gauged away'', so that the parallel transporters
along time links are trivial. This is familiar in gauge theory
under the name ``$A_0=0$ gauge''
 \cite{Henneaux}. 

By definition, a drama is a \system . Its consideration {\em converts function
into structure}. Dynamical laws constrain the structure of the drama. 

Now we are ready to consider deterministic dynamics.
Stochastic dynamics operates in the same way except that enzymes operate with 
certain probabilities.  

An initial state $\mathbf{\Sigma}_t$ of an $N$-th order dynamics shall be a 
subsystem $\S_{[t,t+1-N]}$ of a drama which is generated by subsystems 
 $\S_t,... ,\S_{t+1-N}$ and the time links between them, and enzymes that 
are attached to the objects and links of this \system . The enzymes code
for the constraints on the local structure of the drama which 
determine $\S_{t+1}$ in terms of $\S_{[t,t+1-N]}$. The enzymes must
\begin{enumerate}
\item
determine the descendents in $\S_{t+1}$,
 possible identity links between them
and the existence of time links to and from their ancestors. 
\item
determine non-cognitive links in $\S_{t+1}$. 
They are arrows in $ \S_{[t,t+1-N]}$
composed with time links between $\S_t$ and $\S_{t+1}$
\item 
determine possible cognitive links between objects in $\S_{t+1}$
which are descendents of non atomic objects in $\S_t$. 
\item put enzymes on objects and links of $\S_t$. 
\end{enumerate}

Enzymes of {\em motion} make exactly one descendent of every object, 
with a pair of time-links between it and its ancestor. They only make
 non-cognitive links.

A {\em growth} enzyme makes two or more descendents of one object or,
in case of fusion, makes one common descendent of several objects 
which are linked by identity links. 

{\em death} follows from absence of enzymes that make appropriate objects and 
links.  

I pause to explain the notion of {\em cognitive links} which 
link two non-atomic objects.
By definition, they are \system s  $\X_1$, $\X_2$. 
A cognitive link is a local functor $f: \X_1 \mapsto \X_2$. 
Links between isomorphic \system s are the prototypical examples. 
If there is at most one link between two objects in $\X_1$, then the 
functor is determined by the images $f(Y) $ of objects $Y$ in $\X_1$. 
We use special links -e.g. identity links -
 to connect $Y$ and $f(Y)$. These identity links 
are supposed to be determined by enzymatic computation, i.e. by a 
dynamical process in a \system  \ which is generated by $\X_1, \X_2$
and an initial tentative identity link between some constituents of these.   
The making of cognitive links only makes sense if the dynamics determines
future \system s up to {\em strong} isomorphism, because the internal 
structure of non-atomic objects matters.

\begin{figure}
\begin{center}
\setlength{\unitlength}{0.6mm}
\begin{picture}(120,120)(-10,0)


\put(90,120){$t$}

\put(20,100){\begin{picture}(0,0)
\put(20,20){\circle*{2}}
\put(-5,14){\circle*{2}}        
\put(5,25){\circle*{2}}        
\put(18,19.5){\vector(-4,-1){21}}
\put(5,25){\vector(3,-1){13}}
\end{picture}}

\put(40,120){\begin{picture}(0,0)
\put(0,0){\circle*{2}}
\put(20,0){\circle*{2}}
\put(2,-0.5){\vector(1,0){16}}
\put(18,0.5){\vector(-1,0){16}}
\end{picture}}

\put(60,120){\begin{picture}(0,0)
\put(0,0){\circle*{2}}
\put(20,0){\circle*{2}}
\put(2,-0.5){\vector(1,0){16}}
\put(18,0.5){\vector(-1,0){16}}
\end{picture}}

\put(-5,114){\begin{picture}(0,0)
\put(0,0){\circle*{2}}
\put(20,0){\circle*{2}}
\put(2,-0.5){\vector(1,0){16}}
\put(18,0.5){\vector(-1,0){16}}
\end{picture}}

\put(5,125){\begin{picture}(0,0)
\put(0,0){\circle*{2}}
\put(20,0){\circle*{2}}
\put(2,-0.5){\vector(1,0){16}}
\put(18,0.5){\vector(-1,0){16}}
\end{picture}}  

\put(90,85){$t+\frac 14$}

\put(20,60){\begin{picture}(0,0)
\put(20,20){\circle*{2}}
\put(-5,14){\circle*{2}}        
\put(5,25){\circle*{2}}        
\put(18,19.5){\vector(-4,-1){21}}
\put(5,25){\vector(3,-1){13}}
\end{picture}}
 
\put(40,80){\begin{picture}(0,0)
\put(0,0){\circle*{2}}
\put(20,0){\circle*{2}}
\put(2,-0.5){\vector(1,0){16}}
\put(18,0.5){\vector(-1,0){16}}
\end{picture}}

\put(60,80){\begin{picture}(0,0)
\put(0,0){\circle*{2}}
\put(20,0){\circle*{2}}
\put(2,-0.5){\vector(1,0){16}}
\put(18,0.5){\vector(-1,0){16}}
\end{picture}}

\put(-5,74){\begin{picture}(0,0)
\put(0,0){\circle*{2}}
\put(20,0){\circle*{2}}
\put(2,-0.5){\vector(1,0){16}}
\put(18,0.5){\vector(-1,0){16}}
\end{picture}}

\put(5,85){\begin{picture}(0,0)
\put(0,0){\circle*{2}}
\put(20,0){\circle*{2}}
\put(2,-0.5){\vector(1,0){16}}
\put(18,0.5){\vector(-1,0){16}}
\end{picture}}
\put(40,80){\begin{picture}(0,0)
\put(0,0){\circle*{2}}
\put(-5,14){\circle*{2}}        
\put(5,25){\circle*{2}}        
\put(5.5,22){\vector(-1,-4){5}}
\put(-1,2){\vector(-1,3){3.3}}
\end{picture}}

\put(60,80){\begin{picture}(0,0)
\put(0,0){\circle*{2}}
\put(0,20){\circle*{2}}
\put(-0.5,2){\vector(0,1){16}}
\put(0.5,18){\vector(0,-1){16}}
\end{picture}}

\put(80,80){\begin{picture}(0,0)
\put(0,0){\circle*{2}}
\put(0,20){\circle*{2}}
\put(-0.5,2){\vector(0,1){16}}
\put(0.5,18){\vector(0,-1){16}}
\end{picture}}

\put(25,85){\begin{picture}(0,0)
\put(0,0){\circle*{2}}
\put(0,20){\circle*{2}}
\put(-0.5,2){\vector(0,1){16}}
\put(0.5,18){\vector(0,-1){16}}
\end{picture}}

\put(15,74){\begin{picture}(0,0)
\put(0,0){\circle*{2}}
\put(0,20){\circle*{2}}
\put(-0.5,2){\vector(0,1){16}}
\put(0.5,18){\vector(0,-1){16}}
\end{picture}}

\put(5,85){\begin{picture}(0,0)
\put(0,0){\circle*{2}}
\put(0,20){\circle*{2}}
\put(-0.5,2){\vector(0,1){16}}
\put(0.5,18){\vector(0,-1){16}}
\end{picture}}

\put(-5,74){\begin{picture}(0,0)
\put(0,0){\circle*{2}}
\put(0,20){\circle*{2}}
\put(-0.5,2){\vector(0,1){16}}
\put(0.5,18){\vector(0,-1){16}}
\end{picture}}


\put(90,50){$t+\frac 12$}
\put(20,20){\begin{picture}(0,0)
\put(20,20){\circle*{2}}
\put(-5,14){\circle*{2}}        
\put(5,25){\circle*{2}}        
\put(18,19.5){\vector(-4,-1){21}}
\put(5,25){\vector(3,-1){13}}
\end{picture}}

\put(40,40){\begin{picture}(0,0)
\put(0,0){\circle*{2}}
\put(20,0){\circle*{2}}
\put(2,-0.5){\vector(1,0){16}}
\put(18,0.5){\vector(-1,0){16}}
\end{picture}}

\put(60,40){\begin{picture}(0,0)
\put(0,0){\circle*{2}}
\put(20,0){\circle*{2}}
\put(2,-0.5){\vector(1,0){16}}
\put(18,0.5){\vector(-1,0){16}}
\end{picture}}

\put(-5,34){\begin{picture}(0,0)
\put(0,0){\circle*{2}}
\put(20,0){\circle*{2}}
\put(2,-0.5){\vector(1,0){16}}
\put(18,0.5){\vector(-1,0){16}}
\end{picture}}

\put(5,45){\begin{picture}(0,0)
\put(0,0){\circle*{2}}
\put(20,0){\circle*{2}}
\put(2,-0.5){\vector(1,0){16}}
\put(18,0.5){\vector(-1,0){16}}
\end{picture}}
\put(40,40){\begin{picture}(0,0)
\put(0,0){\circle*{2}}
\put(-5,14){\circle*{2}}        
\put(5,25){\circle*{2}}        
\put(5.5,22){\vector(-1,-4){5}}
\put(-1,2){\vector(-1,3){3.3}}
\put(-1,2){\vector(-1,3){3.3}}\end{picture}}

\put(60,40){\begin{picture}(0,0)
\put(0,0){\circle*{2}}
\put(0,20){\circle*{2}}
\put(-0.5,2){\vector(0,1){16}}
\put(0.5,18){\vector(0,-1){16}}
\end{picture}}

\put(80,40){\begin{picture}(0,0)
\put(0,0){\circle*{2}}
\put(0,20){\circle*{2}}
\put(-0.5,2){\vector(0,1){16}}
\put(0.5,18){\vector(0,-1){16}}
\end{picture}}

\put(25,45){\begin{picture}(0,0)
\put(0,0){\circle*{2}}
\put(0,20){\circle*{2}}
\put(-0.5,2){\vector(0,1){16}}
\put(0.5,18){\vector(0,-1){16}}
\end{picture}}

\put(15,34){\begin{picture}(0,0)
\put(0,0){\circle*{2}}
\put(0,20){\circle*{2}}
\put(-0.5,2){\vector(0,1){16}}
\put(0.5,18){\vector(0,-1){16}}
\end{picture}}

\put(5,45){\begin{picture}(0,0)
\put(0,0){\circle*{2}}
\put(0,20){\circle*{2}}
\put(-0.5,2){\vector(0,1){16}}
\put(0.5,18){\vector(0,-1){16}}
\end{picture}}

\put(-5,34){\begin{picture}(0,0)
\put(0,0){\circle*{2}}
\put(0,20){\circle*{2}}
\put(-0.5,2){\vector(0,1){16}}
\put(0.5,18){\vector(0,-1){16}}
\end{picture}}

\put(60,40){\begin{picture}(0,0)
\put(2,2){\vector(1,1){16}}
\end{picture}}

\put(80,40){\begin{picture}(0,0)
\put(-2,2){\vector(-1,1){16}}
\end{picture}}
\put(5,45){\begin{picture}(0,0)
\put(2,2){\vector(1,1){16}}
\end{picture}}

\put(25,45){\begin{picture}(0,0)
\put(-2,2){\vector(-1,1){16}}
\end{picture}}

\put(40,40){\begin{picture}(0,0)
\put(2,2){\vector(1,1){16}}
\end{picture}}

\put(60,40){\begin{picture}(0,0)
\put(-2,1){\vector(-3,2){20.5}}
\end{picture}}

\put(15,34){\begin{picture}(0,0)
\put(-2,2){\vector(-1,1){16}}
\end{picture}}

\put(-5,34){\begin{picture}(0,0)
\put(2,2){\vector(1,1){16}}
\end{picture}}

\put(35,54){\begin{picture}(0,0)
\put(-18,-18){\vector(1,1){16}} 
\end{picture}}

\put(40,40){\begin{picture}(0,0)
\put(-2,1.2){\vector(-1,2){11}}
\end{picture}}



\put(90,20){$t+1$}
\put(20,-20){\begin{picture}(0,0)
\put(20,20){\circle*{2}}
\put(-5,14){\circle*{2}}        
\put(5,25){\circle*{2}}        
\put(18,19.5){\vector(-4,-1){21}}
\put(5,25){\vector(3,-1){13}}
\end{picture}}

\put(40,0){\begin{picture}(0,0)
\put(0,0){\circle*{2}}
\put(20,0){\circle*{2}}
\put(2,-0.5){\vector(1,0){16}}
\put(18,0.5){\vector(-1,0){16}}
\end{picture}}

\put(60,0){\begin{picture}(0,0)
\put(0,0){\circle*{2}}
\put(20,0){\circle*{2}}
\put(2,-0.5){\vector(1,0){16}}
\put(18,0.5){\vector(-1,0){16}}
\end{picture}}

\put(-5,-6){\begin{picture}(0,0)
\put(0,0){\circle*{2}}
\put(20,0){\circle*{2}}
\put(2,-0.5){\vector(1,0){16}}
\put(18,0.5){\vector(-1,0){16}}
\end{picture}}

\put(5,5){\begin{picture}(0,0)
\put(0,0){\circle*{2}}
\put(20,0){\circle*{2}}
\put(2,-0.5){\vector(1,0){16}}
\put(18,0.5){\vector(-1,0){16}}
\end{picture}}
\put(40,0){\begin{picture}(0,0)
\put(0,0){\circle*{2}}
\put(-5,14){\circle*{2}}        
\put(5,25){\circle*{2}}        
\put(5.5,22){\vector(-1,-4){5}}
\put(-1,2){\vector(-1,3){3.3}}
\end{picture}}

\put(60,0){\begin{picture}(0,0)
\put(0,0){\circle*{2}}
\put(0,20){\circle*{2}}
\put(-0.5,2){\vector(0,1){16}}
\put(0.5,18){\vector(0,-1){16}}
\end{picture}}

\put(80,0){\begin{picture}(0,0)
\put(0,0){\circle*{2}}
\put(0,20){\circle*{2}}
\put(-0.5,2){\vector(0,1){16}}
\put(0.5,18){\vector(0,-1){16}}
\end{picture}}

\put(25,5){\begin{picture}(0,0)
\put(0,0){\circle*{2}}
\put(0,20){\circle*{2}}
\put(-0.5,2){\vector(0,1){16}}
\put(0.5,18){\vector(0,-1){16}}
\end{picture}}

\put(15,-6){\begin{picture}(0,0)
\put(0,0){\circle*{2}}
\put(0,20){\circle*{2}}
\put(-0.5,2){\vector(0,1){16}}
\put(0.5,18){\vector(0,-1){16}}
\end{picture}}

\put(5,5){\begin{picture}(0,0)
\put(0,0){\circle*{2}}
\put(0,20){\circle*{2}}
\put(-0.5,2){\vector(0,1){16}}
\put(0.5,18){\vector(0,-1){16}}
\end{picture}}

\put(-5,-6){\begin{picture}(0,0)
\put(0,0){\circle*{2}}
\put(0,20){\circle*{2}}
\put(-0.5,2){\vector(0,1){16}}
\put(0.5,18){\vector(0,-1){16}}
\end{picture}}

\put(40,0){\begin{picture}(0,0)
\put(20,20){\circle*{2}}
\put(-5,14){\circle*{2}}        
\put(5,25){\circle*{2}}        
\put(18,19.5){\vector(-4,-1){21}}
\put(5,25){\vector(3,-1){13}}
\end{picture}}

\put(60,20){\begin{picture}(0,0)
\put(0,0){\circle*{2}}
\put(20,0){\circle*{2}}
\put(2,-0.5){\vector(1,0){16}}
\put(18,0.5){\vector(-1,0){16}}
\end{picture}}

\put(-5,14){\begin{picture}(0,0)
\put(0,0){\circle*{2}}
\put(20,0){\circle*{2}}
\put(2,-0.5){\vector(1,0){16}}
\put(18,0.5){\vector(-1,0){16}}
\end{picture}}

\put(15,14){\begin{picture}(0,0)
\put(0,0){\circle*{2}}
\put(20,0){\circle*{2}}
\put(2,-0.5){\vector(1,0){16}}
\put(18,0.5){\vector(-1,0){16}}
\end{picture}}

\put(5,25){\begin{picture}(0,0)
\put(0,0){\circle*{2}}
\put(20,0){\circle*{2}}
\put(2,-0.5){\vector(1,0){16}}
\put(18,0.5){\vector(-1,0){16}}
\end{picture}}

\put(25,25){\begin{picture}(0,0)
\put(0,0){\circle*{2}}
\put(20,0){\circle*{2}}
\put(2,-0.5){\vector(1,0){16}}
\put(18,0.5){\vector(-1,0){16}}
\end{picture}}


\end{picture} 
\end{center}
\caption{splitFork dynamics, concurrent version. The time $t+\frac 34$ step is not shown}
\label{fig:concurrent}
\end{figure}
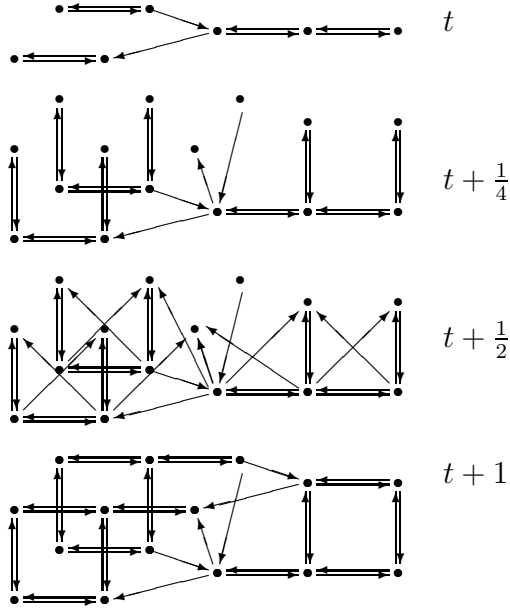

For illustration of the relation between drama and dynamics,
 consider  the splitFork-dynamics. The $t+1$ piece of Figure
\ref{fig:concurrent} shows a portion of a drama. 

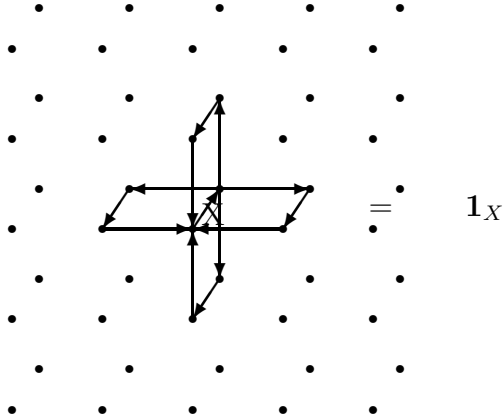
\begin{figure}
\begin{center}
\setlength{\unitlength}{0.6mm} 
\begin{picture}(130,65)(-17,10) 
\thicklines
\multiput(0,0)(0,20){5}{\multiput(10,20)(20,0){5}{\circle*{2}}}
\multiput(6,9)(0,20){5}{\multiput(10,20)(20,0){5}{\circle*{2}}}

\put(30,60){\vector(1,0){20}}
\put(70,60){\vector(-1,0){20}}
\put(50,60){\vector(2,3){6}}
\put(50,80){\vector(0,-1){20}}
\put(50,40){\vector(0,1){20}}

\put(56,69){\vector(-1,0){20}}
\put(56,69){\vector(1,0){20}}
\put(56,69){\vector(0,1){20}}
\put(56,69){\vector(0,-1){20}}

\put(56,89){\vector(-2,-3){6}}
\put(56,49){\vector(-2,-3){6}}
\put(36,69){\vector(-2,-3){6}}
\put(76,69){\vector(-2,-3){6}}

\put(52,61){X}
\put(89,63){= $\qquad \iota_X$}

\end{picture}
\end{center}
\caption{Maxwell Drama. For the Gauss-constraint, the direction to the back is the time direction. For the equations of motion, the upward direction is the time direction. The equations say that the parallel transporter along the path around all the displayed plaquettes  equals the identity. Their is one equation for every pair of links. Formally, the Yang Mills equations are of the same form. Only the gauge group is different}\label{fig:Maxwell}
\end{figure}

Another example 
are 
the discretized Maxwell equations 
on a cubic lattice and in discrete time. This dynamics is of second 
order, so the dynamical laws will involve links in 3 layers 
$\S_{t+1}, \S_t $ and $\S_{t-1}$ and time links. In addition there are
constraints on the initial state, they involve links in $\S_t $ and
 $\S_{t-1}$ and time links between. All these constraints
on the structure of the drama  have the  form
\be
l = \iota_X 
\ee
where $l$ are arrows $X\mapsto X$ that are made from closed paths with the 
above links. They are shown in figure \ref{fig:Maxwell}. 

We explain below why these are the Maxwell equations
including the Gauss constraint.

Let us show that the dynamics is well defined. 
 The links are unitary and the dynamical laws
involve closed paths $l$ with exactly one link $b$ in $\S_{t+1}$.
 Therefore they have  a unique solution 
\be b = u^\ast \ \ \mbox{ if } l = b\circ u  \label{motion:eq} \ee
where $u$ is composed of the remaining links in the path $p$. 
The time links in path $u$ are gauged away and the remaining ones are 
determined by the initial state.

We see that the time evolution merely amounts to promoting arrows $u^\ast $
of the system  to the status of link, while the old links may lose that status but remain as a arrows. In other words, the category  $Cat (\SigmA_t)$  
does not change at all. Recall that this is always the case in motion.

The Gauss constraint is preserved in time. 
This will be shown in section \ref{sec:NCDG} using tools of 
non-commutative differential calculus. 

The gauge group 
${\bf R}$ of electrodynamics 
is determined by the initial state; it is also counted as a constraint on the
initial state.   The coordinatization of links by real vector 
potentials (below) comes from the gauge group ${\bf R}$
similarly as in lattice gauge theory. 
 This can be deduced from the
representation theorems to be proven in section \ref{sec:repres}. 

Therefore real values are attached to the links. If the lattice gauge field 
comes from a vector potential ${\bf A}$ in the continuum they are 
\ba \bullet \rightarrow \bullet &=& 
 \int {\bf A}d{\bf x} , \qquad \mbox{whence} \\
{\Box} &=& \oint_{\Box } {\bf A}d {\bf x} =
 \int_{F:\partial F=\Box} {\bf B }d{\bf f}\ea
for loops around squares. 
${\bf B}$ is the magnetic field.
Parallel squares are surrounded with opposite
orientation. Therefore the total contribution of paths around all space-like 
plaquettes going through $\rightarrow $ is  $\approx$   
 $\mathbf{\nabla}\times {\bf B}\cdot \mbox{area}$, while a time-like plaquette
 gives something proportional to the electric field, since 
$-{\bf E}=\dot {\bf A}$. Putting everything together we get 
Maxwell's equation $\dot{{\bf E}}= \mathbf{\nabla}\times {\bf B}$,
and Gauss' law
$\mathbf{\nabla}{\bf E}=0$. The other Maxwell equation follows from the
 existence of a 4-vector potential.  

Charged fields can be put in \cite{mack:einstein}. 
Note that charge conservation 
is required by the internal consistency of the Maxwell equations -
indestructibility of charged matter is built into a structural description,
it need not be postulated separately.

The Yang Mills equations of elementary particle physics have the same form,
at least formally. Only the gauge group is different. Higgs physics 
can also be put in, at a prize \cite{mack:faddeev}. The world is regarded 
as two sheeted, one sheet carries the left handed matter and the other the
 right handed matter. The two 4-dimensional sheets might be boundaries of  
a five (or higher)-dimensional world. The Higgs fields are possibly non-unitary
parallel transporters between the sheets, as in the model of 
Connes and Lott \cite{ConnesLott},
 but with conventional locality requirements, cf. section
\ref{sec:NCDG}. The prize is that each sheet should have its own 
gauge group $G_L$ resp. $G_R$. The strong gauge group is therefore
$SU(3)\times SU(3)$, broken spontaneously by a Higgs to $SU(3)$. But 
this breaking is expected to produce a massive vector meson, the axigluon
\cite{axigluon} which has not been found until now.  

\subsection{Concurrency}
%

This section presents details on how a well defined dynamics can be composed
from enzymes. There is a technical problem which is known to computer
 scientists as the concurrency problem.
Enzymes specify local structural 
transformations, but the action 
of such transformations at neighboring locations may not commute.

The drama point of view amounts to solving the concurrency problem by a 
generalization
of what is known in applied mathematics as a Jacobi sweep
(as opposed to Gauss Seidel sweeps)\cite{MGrid}. In a Jacobi 
sweep one updates variables attached to nodes and links of a grid
by visiting them one at a time and determining the values of 
their particular variables   at time $t+1$ in terms of time $t$ values of 
variables attached to objects and links in the neighborhood. The result 
is independent of the order of visits.

For simplicity consider first order dynamics, and ignore 
the possibility of making cognitive links. Given $\S_t$,
 the objects of $\S_{t+1}$ need to be made as descendents,
 and the links in $\S_{t+1}$ need to be made.

Divide the time interval in four and let the production of descendents take
place at time $t+\frac 14$ and the initial production of new links at time
$t+\frac 12$. Suppose that the production of descendents and their time links 
is governed by special enzymes called O-enzymes, which are attached to
objects or identity arrows $i$ 
between indistinguishable objects. An O-enzyme may also attach specific
enzymes to the descendents or identity arrow between them. 

There are two types, O2-enzymes attached to 
objects,
and O1-enzymes. 
 The objects $X_i$ connected by identity
arrows with attached O1-enzymes
  form clusters which are in one 1-neighborhood. 
Let us regard the O1-enzymes in this cluster as one O-enzyme. 
It makes a copy of some representative  object $X_i$ in the cluster 
 and links it to all the 
$X_j$ in the cluster by bidirectional time links. 
O2-enzymes at  X
make two descendents  of $X$ and a time link {\em to} one copy and another 
{\em from} the other copy.
This produces a ``fork in time direction''. 
The enzyme may put a bidirectional identity link between the two 
descendents at the ends of the fork prongs.
 Enzymes may be attached to the descendent's identity arrows 
in a manner determined by the O1, O2-enzymes. 

We note that the action of all the O-enzymes anywhere commutes, and so their 
action specifies a globally well defined transformation of $\S_t$ into
$\S_t^\prime$ which consists in the growth of additional elements.
At this stage, $\S_t^\prime $ is only defined as a graph. We make it into a 
\system \ by  extending the composition rule. This is done 
 by specifying the equivalence
 relations involving new links as follows

1. time links are unitary

2. triangles made from time links and identity links are $\iota$.  

 Next we consider L-enzymes. They make links. They act at time $t+\frac 12$. 
Like O-enzymes they are attached to links or 
objects
 in $\S_t$. Their action also consists in the growth of new elements,
 ``diagonal links''. 
 Their action at 
time $t+\frac 12$ makes collections of ``diagonal'' links $l$.
They  connect objects in $\S_t$ with
descendents of objects in $\S_t$ in a manner which depends on the enzyme, 
on the
link or 
object
 it is attached to,  on 
the neighborhood of this link or object in $\S_t$,
 and  on the descendents of objects in this neighborhood. 
It may also attach enzymes to the newly made (diagonal) links. 
The new links $l$ are made by composition of links in $\S_t$ and time links.
Furthermore, the L-enzymes  may put marks on the new links which 
will serve as indicators of  adjoint relationships to be specified later. 
We note that the action of all the $L$-enzymes
anywhere commutes, since links are created depending only on what was before. 
Therefore the action of all L-enzymes at time $t+\frac 12$ 
specifies a globally well defined transformation 
of $\S_t^\prime$ into $\S_t^{\prime\prime}$ which consists
in the growth of new elements. $\S_t^{\prime\prime}$ is a \system \ because the
new links are arrows of $\S_t^\prime  $. 

At time $t+\frac 34$ we lift the ends (sources or targets) of diagonal links 
which are in $\S_t$ by composing them with time links to or from descendents.
The time links were made such  that the way  to do this is unique.
There results a well defined \system \ $\S_t^{\prime\prime\prime}$ with 
$Cat(\S_t^{\prime\prime\prime} )$ = $Cat(\S_t^{\prime\prime} )$.

Finally, at time $t+1$, one considers pairs  $Y_1,Y_2$ of of descendents 
arbitrary objects such that there is some link between them. The local action
at $(Y_1,Y_2)$ consists in an examination of the totality $T$ of links between
 $Y_1$ and $Y_2$ and in declaring some of them  adjoints of others
 in a manner which depends on $T$ and the marks on the links. 
How this is done must be specified 
a priori by specifying the meaning of marks. We note that the local action 
for different pairs commutes. Therefore there results a well defined 
\system \ $\S_t^{\prime\prime\prime\prime} $ with 
$Cat(\S_t^{\prime\prime\prime\prime} ) = Cat(\S_t^{\prime\prime\prime} )$.

The objects which are descendents of objects in $\S_t$ and the links between 
them generate $\S_{t+1}$

This demonstrates validity of the following 
\begin{theorem}{\em (Deterministic 1. order enzymatic dynamics)}
Suppose that $\S_t$ can be obtained from a \system \ without attached enzymes 
by attaching O-enzymes and L-enzymes (as described above) to links and
objects.
 Then the enzymes determine a 
unique map $\S_t \mapsto \S_{t+1}$
to another \system . If $\S_t $ is the time $t$-layer of the part 
$\S_{\leq t}$ of a drama, then  $\S_{\leq (t+1)}$ is defined.  
\end{theorem}

The making of cognitive links requires a separate consideration
 in order to fix the extension of the composition law to them.

Given a collections $f_{21}$ of identity links from objects of \system \
$\X_1$ to \system \ $\X_2$, and $f_{32}$ from $\X_2$ to $\X_3$, this 
defines a possibly empty collection $f_{31}=f_{32}\circ f_{12}$.
If $f_{21}$ and $f_{31}$ define local functors, then so does $f_{31}.$
By definition, cognitive links are functors.
We define arrows as equivalence classes of paths. Equivalences are generated 
by equivalences of paths without cognitive links, and equivalence of
two paths $(b_1,...,b_n)$ when all $b_i$ are cognitive links 
and their composition defines the same local functor for the both paths.

\section{Transformation theory}
\label{sec:repres}
This section will show in what way \system 's theory is a generalization
of gauge theory. Some basic concepts and tools flow from this. 

In quantum mechanics, Dirac's transformation theory played 
an important role \cite{Dirac}.
 It rests on the fact that unitarily equivalent 
representations of the  algebra ${\bf A}$ of observables in Hilbert spaces
are not physically distinct. The spectral theorem assert the existence 
of  representations in which  given commuting observables are
simultaneously diagonal, i.e. act as multiplication operators on 
function spaces. 

Similarly, isomorphic \system s
are not considered distinct. Here I present representation theorems 
and some properties of special representations are pointed out. 

I explain the notion of a {\em representation}. 
In group theory, a representation of a group $G$ is not just a homomorphism
(structure preserving map)  to 
another group $G^\prime$, but it is required that $G^\prime $ must come with 
 some predefined structure, and the group operations must
 be compatible with it. More particularly, $G^\prime$ must consist of 
linear maps of a vector space, and group multiplication must be 
composition of maps. 

Similarly, a portrait in oil is a structure preserving map of a person. 
The image is supposed to consist of oil paint on canvas.

Generalizing this, representations of a \system  \ will be 
defined as local functors  to  instances of a class of \system s which
 are equipped with some 
predefined structure, and operations like composition $\circ$ of arrows 
are supposed to be compatible with it.  
Sometimes there is additional structure (e.g. composition rules for objects)
which are required to be preserved. 

A representation is 
called semi-faithful when no two objects or links are mapped into the same 
object or link, faithful when the same is true of arrows. 

There are many kinds of representations. The most important ones have
the following classes of systems as images
 
- communication networks: The arrows $f :X\mapsto Y$  are maps 
 $ {\bf f}: \Omega_X \mapsto \Omega_Y $ of sets 
(or spaces with more structure) and composition is composition of maps.

- archetypes

- unfrustrated systems

\noindent Archetypes are special \system s , often with few objects. 

\begin{theorem}{\em (Representation as a communication network)}. 
\label{rep:1}
Every \system \ with finitely many objects and links admits a faithful 
representation as a communication network, i.e. there are sets $\Omega_X$
associated with objects, and links and arrows are maps 
${\bf f}:\Omega_X \mapsto \Omega_Y$.
\end{theorem}
Given a path ${\cal C}=(b_1,...,b_n)$ from $X$ to $Y$,
 let $f=b_n\circ ...\circ b_1$.
Then ${\bf f}: \Omega_X\mapsto \Omega_Y$ is called the 
{\em parallel transport along the path {\cal C}}.

There is a more elaborate version of theorem \ref{rep:1} wherein there are 
separate
 {\em input spaces} $\Alpha_X$ and {\em output spaces} $\Omega_X$,
 objects $X$ define maps $\iota_X: 
\Alpha_X \mapsto \Omega_Y$ and links and arrows are maps $\Omega_X
\mapsto \Alpha_Y$. The theorem was proven in \cite{mack:einstein}; it could 
be generalized to \system s with more elements. 
The proof is constructive, but the construction of $\Omega_X$ involves 
elements of the whole \system . 
\begin{theorem}{\em (Principal fibre bundle representation)}
\label{theo:princRep}
Let $\S$ be a \system \ with countably many objects and links, and 
suppose that its links
 are all unitary. Then it admits a representation as in 
theorem \ref{rep:1}, where $\Omega_X$ are copies of the gauge group $G$,
and the maps commute with the right action of $G$ on the $\Omega_X$ by
group multiplication.  
\end{theorem}
Maps which commute with the right action of $G$ amount to left 
multiplication with group elements. 

 The theorem
recovers the structure of lattice gauge fields in pure lattice gauge theory
under the single extra structural assumption 
$$ forth \circ back = identity .$$

\begin{corollary}{\em (associated vector bundle representation)}
\label{theo:assocRep}
Under conditions as in theorem \ref{theo:princRep}, if the gauge group $G$
admits  a faithful representation in a vector space $\Omega $, there is
a representation as in theorem \ref{rep:1} where $\Omega_X$ are copies of 
$\Omega$, and the maps ${\bf f}$ are linear.  If the linear representation 
is a unitary representation in a Hilbert space, then the arrows
are unitary maps of Hilbert spaces. 
\end{corollary}
This is the standard construction of associated vector bundles
from principal fibre bundles. 

{\sc Proof of theorem} \ref{theo:princRep}:
The graph of a \system \ with countably many objects and links admits a 
spanning tree which generates an unfrustrated \system \ ${\bf T}$. Let 
$X_0$ be its root and identitfy $G=G_{X_0}$. To every object $X$,
 there is a unique unitary arrow
${\bf T} \ni h_X: X_0 \mapsto X$. Associate copies of the gauge group $G$ 
with objects $X$; they are all identified with $\Omega_{X_0}=G$ via the maps 
$h_X$. Convert arrows $f:X\mapsto Y$ into elements ${\bf f}_G$ of the gauge 
group $G_{X_0}$ 
according to ${\bf f}_G = h_Y^\ast \circ f \circ h_X$. ${\bf f}_G$ acts
on $\Omega_X=G $ by ${\bf f}_G(g) = {\bf f}_G g \in G=\Omega_Y $. 
In this way we construct a 
\system \ which is isomorphic to the original one and has the desired 
properties.

\subsection{Logic}
\label{sec:logic}
\system s with unitary links are generalizations of gauge theories. There is a
quite different class of \system s (and of categories \cite{Lawvere,topos})
 where 
links and arrows are maps of sets which need be neither surjective nor 
injective. Logic belongs here.  

To give an example what can be done with representations, I report
some theorems on logic. They were proven by Schrattenholzer in his 
thesis \cite{schrattenholzer}. 
I will not reproduce the proofs. 
\begin{define}\emph{(Logical archetype)} 
The logical archetype is the \system \ with two objects denoted $T$ 
({\sc true}) and $F$ ({\sc false}), and  
links $e: T \mapsto F,\quad e^{\ast }: F\mapsto T, \qquad o: F\mapsto F$, \\
The composition law is defined by the following relations:\\
$ e\circ e^{\ast } = \iota_F, \ e^{\ast }\circ e = \iota_T,\
o\circ o = o = o^{\ast}, \ o\circ e = e, \quad e^{\ast} \circ o 
= e^{\ast}.$ In addition there is a rule for composing objects with 
the help of the Scheffer stroke $|$ : $T|T=F$, $T|F=F$,
$F|T = F$, $F|F=T$. 
\end{define}
In the logical archetype and in all our logical systems,
the links are interpreted as ``excludes'', and  the special case of 
unitary links as ``not''. A pair of adjoint unitary links is graphically 
represented as 
$\sim $.  The objects represent
 potential propositions, and the Scheffer stroke $|$ is interpreted as 
``neither nor'' (NOR). Hence $A|A$ is interpreted as  $ not \ A$. . 

Note that logically, ($A$ excludes $B$)
 implies ($B$ excludes $A$), since both are 
equivalent to the statement that $A$ and $B$ are not both true. This rule 
of logic says that adjoints of links should be  links. 

In the following, we are relaxing assumption 1 by admitting links from a
composite object to   its constituents  and vice versa. 

Define a {\em logical  \system} as a \system \ without equivalence relations 
between paths other than possible unitarity of links, in which some composite 
objects $A|B$ may appear, subject to the following conditions: 

1. With 
$A|B$ also $A$ and $B$ are in the \system , and there are 
pairs of adjoint  links 
$A \leftrightarrow A|B \leftrightarrow B$. 

2. For every object $A$, including  composite objects $A=B|C$ there is an 
object $A|A$. Furthermore $A$ and $A|A$  are linked by  adjoint
unitary links $\sim$ in both directions.

A logical representation of a (logical) \system \ is a local functor
 into the logical archetype, subject to the additional requirement
\be F(A|B) = F(A)|F(B). \ee
Note that it maps every object of the \system \ into $T$ or $F$. In this 
way truth values are assigned.

\begin{theorem}[Schrattenholzer 1999] 
In a logical representation of a logical \system ,\ truth values 
$T$, $F$ are assigned in accordance with the axioms of proposition logic.

Conversely, let $\S$ be any \system \ (not necessarily a logical one), 
possibly with composite objects $A|B$. If there 
exists an assignment of truth values which is consistent with 
proposition logic (with the 
above interpretation of objects,  links and Scheffer stroke $|$ ), 
then the \system
\ admits a logical representation.
\end{theorem}

The propositional content in a logical \system \ is in its links. 
They are subject to being transformed by the moves shown in 
\ref{fig:logic}. It was proven by Schrattenholzer that these
rules are complete - every deduction of proposition logic is possible with 
their help. Furthermore, he showed that there is also a local decision 
calculus. The diagrams in figure \ref{fig:logic} may be regarded as 
compound arrows if 
two parallel links or arrows $f_1:X\mapsto Y $ and $f_2:X\mapsto Y$ 
can be regarded as a single arrow, cp. later (definition \ref{def:semiadd}). 
\begin{figure}
\unitlength1cm
\label{fig:logic}
\begin{picture}(1,1.4)
\put(0.5,1.2){1) adjoint of a link}
\end{picture}

\begin{picture}(1,1.4)

\put(0.5,1){2) A}
\put(1.4,1.1){\circle*{0.1}}
\put(1.6,1.1){\line(1,0){1}}
\put(2.6,0.95){\framebox(0.5,0.3)}

\multiput(3.1,1.25)(0.1,0.1){5}{.}
\multiput(3.1,0.95)(0.1,-0.1){5}{.}


\put(3.7,1.85){\circle*{0.1}}
\put(3.7,0.35){\circle*{0.1}}

\put(3.8,1.75){\line(1,-1){0.5}}
\put(3.8,0.45){\line(1,1){0.5}}

\put(4.4,1.1){\circle*{0.1}}

\put(4.6,1){B}

\end{picture}

\begin{picture}(1,1.3)

\put(0.5,1){3) A}
\put(1.4,1.1){\circle*{0.1}}
\put(1.6,1.1){\line(1,0){1}}
\put(2.8,1.1){\circle*{0.1}}
\put(3,0.6){\Huge\~}
\put(3.6,1.1){\circle*{0.1}}
\put(3.8,1.1){\line(1,0){1}}
\put(5,1.1){\circle*{0.1}}
\put(5.3,1){B}
\end{picture}

\begin{picture}(1,1.3)

\put(0.5,1){4) A}
\put(1.4,1.1){\circle*{0.1}}
\put(1.5,0.6){\Huge\~}
\put(2,1.1){\circle*{0.1}}
\multiput(2.2,1.1)(0.1,0){10}{.}

\put(3.2,0.95){\framebox(0.5,0.3)}

\multiput(3.7,1.1)(0.1,0){11}{.}
\put(4.9,1.1){\circle*{0.1}}
\put(5,0.6){\Huge\~}
\put(5.5,1.1){\circle*{0.1}}
\put(5.7,1){B}

\put(3.45,1.25){\line(1,1){0.4}}
\put(3.45,1.25){\line(-1,1){0.4}}
\put(3.45,1.65){\oval(0.8,0.8)[t]}
\end{picture}

\begin{picture}(1,1)

\put(0.5,1){5) A}
\put(1.4,1.1){\circle*{0.1}}
\put(1.6,1.1){\line(1,0){1}}
\put(2.6,0.95){\framebox(0.5,0.3)}
\multiput(3.1,1.1)(0.1,0){11}{.}
\put(4.3,1.1){\circle*{0.1}}
\put(4.4,0.6){\Huge\~}
\put(4.9,1.1){\circle*{0.1}}
\put(5.1,1){B}

\put(2.85,0.4){\circle*{0.1}}
\multiput(2.8,0.9)(0,-0.1){6}{.}
\put(1.55,1){\line(2,-1){1.2}}

\end{picture}
\caption{Schrattenholzer moves to make logical deductions. Every one of the above compound arrows between $A$ and $B$ may be replaced by a link pair $A-B$. Lines represent adjoint pairs of links, and $\bullet \dots \Box \dots \bullet $ stands for a composite object $C|D$ with links to its constituents $C, D$; $\mathbf{\widetilde{\quad}}$ is a unitary link pair, interpreted as ``is not''.}
\end{figure}

\subsection{Frustration}
In the context of the representation theorem \ref{rep:1}, 
paths $C = b_1, ... , b_n$ from $X$ to $Y$ define parallel 
transport ${\bf f_C}:\Omega_X \mapsto \Omega_Y$, 
of elements in $\Omega_X$ and frustration exists if 
the parallel transport along different paths does not agree. Local 
frustration occurs when this happens for paths which stay in a neighborhood.
This situation appears many times in physics under different names. 
(If all links are unitary, frustration exists if and only if the gauge group
is nontrivial)

- {\em Curvature} in general relativity and in the Riemannian geometry of
surfaces in 3-dimensional Euklidean space.
Space time is curved in general relativity if the parallel transport
of a tangent vector [e.g. a 4-velocity] from space time point $X$ to $X$
along two different paths need not give the same result. 
Generically, the gauge group is the Lorentz group.

- {\em Field strength}
in electromagnetism and in the gauge theories of elementary particle physics,
where vectors in color spaces are parallel transported.
 
- {\em Arbitrage} in financial markets, and

- {\em Frustration} in 
spin glasses. 

In a spin glass model, one has spins attached to the sites $X$ of a lattice.
They may either point up or down. There are links between some of them. 
They are assigned values $+1$ if it is energetically favorable for the 
spins at their ends to be parallel, and $-1$ if antiparallel is favored. 
There is frustration if the requirements for energetically
favorable alignment of spins are in conflict.
\footnote{This explains the name. In real life there is frustration if  
 one's different desires cannot all be fulfilled because 
 they are mutually incompatible.}  

A gauge theory of financial markets was presented by Ilinski \cite{finance}

Eschers impossible pictures provide other examples of frustration: 
People move around and up and up,
yet arrive back  at their starting point. 
In a
representation of a  three-dimensional scene, this
is impossible, because  change of height should be 
path independent.
 The pictures also illustrate the 
point that frustration prevents the assignment of global meaning (height) by
synchronization. There is a lot of frustration in human communication.
 We do not exclusively communicate facts which have a globally defined meaning.

If we think of the absence of a fundamental adjoint as a kind of 
frustration also, 
\footnote{
$ forth \circ back \neq identity$ in some sense 
when forth is a one way street}
 then in essentially all our examples nontrivial changes
in time are associated with frustration. 
Thus, {\em change is caused by frustration.}

\section{Managing complexity}
The main theme of this section is the management of complexity by
 construction of simplified models, also known as effective theories,
which operate on coarser scales. Thermodynamics and electrodynamics
of polarizable media are well known examples of effective theories in physics. 

Before coming to this I will discuss emergence as a manifestation of 
complexity, and some alternative strategies to deal with complexity. 
%
\subsection{What is complexity, emergence?}
Emergence is sometimes described in terms like these. 
``From several components 
which happen to get together something fundamentally new originates,
 often with totally unexpected properties. The classical example is water whose
properties are not predictable from those of hydrogen and oxygen.''
It is characterized as  surprising behavior which cannot be anticipated form 
the behavior of their {\em isolated} parts 
\cite{Casti,Emmeche,Baas:1}. 
But theoretical chemists 
can predict the properties of water from those of hydrogen and oxygen atoms.
The mystery comes from ignoring the links, here chemical bonds. 

Here I am only interested in emergence in complex systems. 
I consider a system as genuinely complex if it shows behavior which cannot be
understood by considering small subsystems in isolation. 
Such behavior I call emergent. Since they do not show in small subsystems,
such emergent phenomena are nonlocal phenomena. We wish to understand them as a
consequence of local interactions, including those from links between 
constituents of different subsystems of any kind. 

This is not a task which is hopeless by definition. 

In quantum field theory there are several known mechanisms which
lead to emergent phenomena, and string theorists are exploring more \cite{strings}.

In general, there are several strategies

- large scale computer simulation

- special mechanisms 

- exploitation of symmetry

- multiscale analysis

Although it is true that computers get faster much more quickly than 
scientists get smarter, it is not true this will solve all problems in due
 time. Large scale computer simulations do not qualify as a universal 
brute force method because the computer does not tell what to look for. 

\subsection{Special mechanisms}
Special mechanisms of particular interest for life  include 
those discussed in section  \ref{sec:examples},
 including the the splitFork dynamics.
They are based on propagating shock waves.

Several mechanisms are known in gauge field theory which lead to nonlocal
phenomena, besides propagating harmonic waves. Typically they imply 
protection of wave propagation against nonlinear perturbation, forbidding
generation of masses.

1. Gauge invariance in gauge theories whose gauge group possesses a nontrivial
 center $\Gamma$.
\footnote{The center of a group $G$ consists of those elements which 
commute with all elements}
 This includes theories with an Abelian (=commutative)
gauge group   like Maxwell's theory. Gauss' law asserts that 
 the presence of central charges causes flux which can be observed arbitrarily far away.
Central charges come from matter fields which transform non-trivially under 
$\Gamma$. Electrically charged fields in Electrodynamics and quark fields in 
Quantum Chromodynamics are examples. In the latter example, long lines of flux
cost too much energy, therefore quarks are confined and physical states carry no central charge \cite{mack:confinement}. 

2. Chiral invariance. The deeper reason behind this is the Atiyah Singer 
index theorem  applied to the Dirac operator $D$ in an
 external gauge field. It implies that $D$ must have zero modes for certain
 boundary conditions, in numbers depending on them. This sensitivity 
to boundary conditions implies that the Greens function for $D$ - the
 Dirac propagator - will have 
infinite correlation length when it exists at all, for arbitrary gauge field. 
Contrast the covariant Laplacian in an external gauge field with non-vanishing 
field strength. It is strictly positive and its Green's function decays 
exponentially \cite{Balaban:GreenFct}. The Atiyah Singer index theorem 
applies in the continuum. The approximate zero modes have also been 
found in computations of the spectrum of the Kogut Susskind 
discretization of the Dirac operator on a lattice \cite{Kalkreuter}. 

3. Supersymmetry  and additional mechanisms 
now under consideration in string theory cannot be discussed here.

\subsection{Exploitation of symmetry}
Symmetries are properties of a system as a whole which are often rather easy
to detect. When detected, they can be exploited.

A crucial problem in complex systems is often 
the determination of the long distance behavior. In favorable circumstances 
this problem can be solved or reduced to manageable problems by exploitation of
symmetries. Let us consider examples. 

Propagation of waves is an emergent phenomenon because local equations 
of motion lead to nonlocal phenomena. Electromagnetic waves, sound waves,
and matter waves
are known examples. In homogeneous media, translation symmetry can be 
used to reduce the problem to one which is no longer complex in our sense.
Consider for instance the wave function $\Psi ({\bf x})e^{-i\omega t}$ 
of noninteracting electrons in a potential which is invariant under 
lattice translations ${\bf x} \mapsto {\bf x} + \sum n_i {\bf e}_i$,
 $n_i \in {\bf Z}$. Herein, 
${\bf e_i } (i=1,2,3)$ are some given vectors which define a lattice.
One must solve the 1-particle Schr\"odinger equation.
${\bf x}$ could be  points of a continuous space or of a discretization of it.

 One introduces 
Bloch waves which are invariant under lattice translations,
\ba
\Psi (x) &=& e^{i{\bf k x}}u_{\bf k}(x) , \\
u_{\bf k}({\bf x} + {\bf e}_i) &=& u_{\bf k}({\bf x}),
\ea
and one is left with  Schr\"odinger equations 
for $u_{\bf k}$ on a single lattice cell with
periodic boundary conditions, i.e.
 on a compact space without ``large distances''.

A more subtle example is the treatment of statistical mechanical systems 
(in equilibrium) at a critical point. 
By definition, the correlation length is infinite at a critical point;
therefore there are correlations between very distant regions of space,
and so the system is complex in our sense to begin with.
 The problem is to find the 
long distance behavior. Under suitable conditions, the long distance 
behavior can be described by a field theory 
which is invariant under all conformal transformations.
It was shown i the  seventies \cite{mack:conformal}
that any such theory can be regarded as living on a compact space
and there is a ``conformal Hamiltonian'' which has a purely discrete 
spectrum $\geq 0$ with only a finite number of eigenstates below any finite 
value $E$. This is true in any dimension $\geq 2$; in 2 dimensions one needs 
the extra assumption of half integral spin. The assertion about the spectrum 
assumes Wilson operator product expansions as asymptotic expansions
(they are then automatically summable to convergent expansions
 \cite{mack:ope}). On a compact space, there are no large distances any more. 

Another subtle class of ``manageable'' systems are integrable models
\cite{Faddeev:integrable}
Again, their treatment involves subtle transformations to systems which are 
in a sense ``no longer complex''. The author is not prepared to enter
into a  discussion of these methods, although they deserve
mention here because they exploit the property of certain 
equations of motion that can be expressed as a requirement of no frustration. 

Nearly all the standard methods of theoretical physics 
to deal with complex systems are based on the exploitation of
symmetries. Often one regards the system of interest as obtained by
perturbing a ``free'' system. The free system is solved by exploitation of 
symmetries, and the perturbation expansion involves calculations within 
the framework of the free theory. 

However, these methods are limited in their applicability.

\subsection{Multiscale analysis}
\label{sec:multiscale}
The general idea of multiscale analysis is that,
 although by definition complex systems cannot
be understood by examining small subsystems in isolation, a complexity
reduction can be achieved by doing so. It constructs  
objects and links of a new \system \
whose objects represent subsystems of the old one, but which have much fewer
degrees of freedom. The new \system \  is still complex, but typically the 
procedure can be iterated. In practice, few repetitions suffice
because the number of objects decreases exponentially.  

The axiomatic properties of  \system s are not quite suitable for the purpose
of multiscale analysis,
but there is a natural way to extend them without seriously violating the 
philosophical principle of {\em minimal a priori structure}.  

It is natural to admit the possibility that
two (or more) parallel links or arrows $b_1:X\mapsto Y $ and $b_2:X\mapsto Y$ 
are regarded as a single arrow, denoted $b_1 \oplus b_2$. 
I emphasize that {\em no assumption of linearity is involved} at this stage.
In the extreme case, $\oplus$ could be a direct sum.
\begin{define}{\em (Semi-additive System)}.
\label{def:semiadd}
A {\em semi-additive system} $\S^+$ satisfies the axioms of a \system ,
 except that
arrows may be composed from links and their adjoints with the help of
two operations, $\circ$ and $\oplus$. The $\oplus$-operation makes the 
set of all arrows with given source $X$ and target $Y$ into an additive  
semigroup. The distributive law holds
$$ (f_1 \oplus f_2)\circ (g_1 \oplus g_2) = 
(f_1\circ g_1) \oplus  (f_2\circ g_1) \oplus (f_1 \circ g_2) \oplus (f_2 \circ g_2)$$

A local functor $F$ of a \ssystem \ obeys $F(g \oplus h) = F(g)\oplus F(h)$.

An arrow  $o$  is a zero arrow if $f\oplus o = f$ for all $f$.
 It is understood 
that arrows are modulo zero arrows. 
\end{define}
If $f\oplus h = g \oplus h $ implies $f=g$, whatever $h$, then the additive
semigroup admits a unique  extension to an additive group.

 But there are some important examples, 
so called {\em discrete event dynamical systems} (DED's) \cite{DED}
where this property does not hold. They have real (or matrix) valued links 
(similarly as in Maxwell theory), with addition $+$ as composition $\circ $,
 and $f\oplus g = max(f,g)$. In the case with one object, this is 
the {\em time table} or {\em Max Plus\/}-''algebra''. For matrices, the 
maximum is taken entry by entry. 

In the rest of this section, I work in the category of \ssystem s. 
and the word {\em system} shall mean \ssystem.

\subsubsection{Deterministic case. Multigrid methods}
\label{sec:MGrid}
For definiteness sake, I consider iterative solution of optimization problems.
This is a very general class of problems. For instance, finding a solution
$x$ of an equation $f(x)=g$ is equivalent to finding a minimum of 
$dist(f(x),g)$ if $f(x)$ and $g$ are in a metric space with distance $dist$.

We seek $\S$ in a class ${\cal S}$ of systems such that a given local 
cost function $\H $ is minimized.  $\H $ assigns a real number $\H (S)$ to 
every $\S\in {\cal S}$ which is a sum of contributions from
 neighborhoods. One
seeks approximate solutions. Therefore a criterium for a tolerable error 
should also be specified. 

In  the computation of an iterative solution
one starts from a rough approximation $\S=\S_{init}$ and one has 
a collection of local structural transformations (enzymes) to act on 
$\S \in {\cal S}$.  I assume that it contains enzymes 

1. to solve the local problems

2. to compose links with $\circ$ and $\oplus $.

The precise local problems depend on the problem, but they are always 
optimization problems for subsystems of individual neighborhoods within $\S$.
Put another way we wish to reduce a global optimization problem to a local one.

Under the stated assumptions, there is a universal problem solving strategy,
{\em relaxation}. One sweeps through the system and determines what is 
variable in individual links and objects in such a way that the local cost 
function is minimized subject to the constraint that everything else remains
constant. Because of locality of the cost functional, this is a local problem. 

When relaxation is very slow to converge, one speaks of {\em critical slowing down}. It is a typical effect of genuine complexity, because relaxation is 
inefficient in dealing with nonlocal phenomena. Multiscale analysis comes 
in when there is critical slowing down. The problem that one may get stuck in
local minima is something else again; it is not the issue under consideration 
here. 

In a multiscale analysis one introduces levels 0,1,2,... .
 The system $\S$ is level 0. One constructs further systems
 $\S^1, \S^2  , ...$
called level 1,2,... and links between them. In a genuine multigrid
(as opposed to unigrid) the only links between levels connect
$\S^j$ and $\S^{j+1}$. 

An object $X^1\in \S^1$ represents a subsystem of $\S$, and similarly for 
links. $\S^1$ varies with $\S $ . It is to be constructed together with a
 local cost function ${\H }^1(\S^1) $  in such a way that

1) The conditional minimum $\tilde{\S} (\S^1)$ of $\H (\S )$ under the
 constraint that $\S^1 $ is fixed can be found by fast converging relaxation 
 in level 0. $\tilde{\S} (\S^1 )$ will be called the optimal interpolation of 
 $\S^1$.

2) If $\S^1$ is a minimum of ${\H}^1$
 then $\tilde{\S}(\S^1)$ is an approximate 
 minimum of $\H $ which can be corrected by further relaxation sweeps on 
level 0. 

In this way the problem is reduced to minimization on level 1. To solve this
problem, one introduces level 2, and so on. The technical aspects of how to 
organize the whole iteration scheme (V-cycles, W-cycles ... ) shall not 
interest us here \cite{MGrid}. 

The objects and links of $\S^1$ may contain data which reflect the structure
of the corresponding subsystems of $\S$, but they do not determine it 
uniquely. In this sense there is complexity reduction. Typically, 
the number of elements (links and objects) in level 1 is only a fraction of 
those in level 0, and the numbers of internal degrees of freedom of individual 
elements is about the same. The idea is that one only retains structural 
information to the extent that it is relevant for the cooperation of 
subsystems as a whole that is responsible for nonlocal emergent phenomena. 

The problem with this method is the fulfillment of the above requirements 1)
and 2). There are very different kinds of optimization problems, 
and only a part of them can be successfully treated with existing multigrid
technology. Deterministic equations of motion are intractable except in
favorable cases.


Let me describe an 
\begin{example}{\em (discretized linear elliptic PDE's)}
\label{example:multigrid}
We seek the solution of a system of linear equations
for vector valued functions $u=\{ u_z\}$, 
\be \sum_w L_{zw} u_w + f_z = 0 \ .  \label{Lu=f}\ee
or $Lu+f=0$ for short. 
$L_{zw}$ are linear maps (matrices), and $f$ is a given vector function
(section, really ...), $u_z, f_z \in \Omega_z $.
\end{example}
We assume that $L$ is a positive operator. 
The system theoretic interpretation is as follows. 

I work within the context of the 
associated vector bundle representation theorem, corollary \ref{theo:assocRep},
with Hilbert spaces $\Omega_z$ with scalar product $<.>$. 
Addition $\oplus $ of links shall  be written as $+$, 
symbols $\circ $ are omitted, and $0$ is the zero link.  

$\S$ shall consist of a constant {\system} $\bar{\S} $ 
with at most one link $L_{wz}:z\mapsto w$ for every pair $z, w$ of objects,
plus one object $\infty $ 
with associated vector space $\Omega_\infty$. For simplicity, admit and take
$\Omega_\infty = {\bf R}$,
possibly not isomorphic with $\Omega_z, z\in \bar{\S}$, and identify 
$f_z, u_z$ with maps ${\bf R} \mapsto \Omega_z$, viz. $r \mapsto u_zr$. 
In this way, $f_z$ and $u_z$ become links $\infty \mapsto z$. 

For every object
$z\in \bar{\S}$, there shall be a constant link 
(linear map) $f_z: \Omega_\infty \mapsto \Omega_z$ and one variable link 
(linear map)  $u_z: \Omega_\infty \mapsto \Omega_z$. The cost functional 
shall be quadratic, 
\be {\H} ({\bf S}) \equiv E(u) = \frac 12 \sum_{z,w} <u_w , L_{wz} u_z> + 
\sum_z <f_z, u_z> \ , \nn \ee     
The local problem is the solution of some equation $L_{zz}u_z+g_z=0$
for individual $z$. Write
its solution as $(-L_{zz}^{-1})g_z$, without implying that negative and 
inverse can be computed separately.
$(-L_{zz})^{-1}$ are loops.  Relaxation at $z$ updates
$$ u_z \mapsto (-L_{zz}^{-1})[u_z + \sum_{w\neq z} L_{zw} u_w] .  $$
This only involves solution of the aforementioned local problem, and
composition and addition of links.  By assumption, there are enzymes to solve
it. The iterative solution is 
\be u_z = (-L_{zz})^{-1} \sum_w \sum_{p: w\mapsto z} {\bf p} f_w
\label{relax}
\ee
where the sum is over all paths $p$ with links $L_{zw}(-L_{ww}^{-1})$,
and ${\bf p}$ is the arrow associated with the path $p$. 
This illustrates once again the principle of universality and minimal a priori 
structure. The axiomatic composition operations $\oplus$ and $\circ$ suffice
to reduce a global to a local problem, and if the local problem is
a global one on a finer scale, the procedure can be repeated.

If there are 
important contributions in (\ref{relax}) from very long paths,
 the iteration is slow to 
converge, and there is critical slowing down. This happens when $L$ has very
 small eigenvalues. The multigrid method deals with this. For reasons of
space, I will not give details here but treat the stochastic case instead. 


\subsubsection{Stochastic case. Renormalization group}  
\label{sec:RG}
The setup in the stochastic case is as in the deterministic case
except that now we don't want to minimize a cost function $\H$, but 
study probability distributions for systems $\S\in {\cal S}$ of the form
$p(\S)=Z^{-1}e^{-\beta \H(\S )}p_0(\S)$ where the a priori distribution 
$p_0$ assigns equal probabilities in some sense.
We are also interested in properties of the expectation values. 
Thus, $\S $ are random systems, and their elements are random variables.

In a critical situation, we expect long range correlations. 
Long range is measured by path length.
If $\S$ is a drama, the long range correlations could be in time.
\footnote{Haken's slave principle \cite{Haken} was invented to deal with this 
case} 

Starting from level 0, one introduces systems $\S^{1} $ of level 1 and 
a probability distribution $p^1(\S^1)$ for them.
 The links and objects of level 1 
represent subsystems of $\S$ in the same way as before, and they vary with
$\S$. One considers the conditional probability distribution 
$p(\S|\S^1)$ for $\S$, given $\S^1$. One demands  that 

1. The conditional probability distribution $p(\S|\S^1)$ for $\S $ 
shows no long range correlations.

2. $p({\bf S})= \sum_{\S^1} p(\S | \S^1)p^1(\S^1).  $

Typically, the objects $x$ of $\S^1$ contain data $\Phi_x $
which reflect the 
structure of the subsystems $\X$ of $\S$ to which $x$ corresponds.
It does so  to the extent
that it is relevant for cooperative effects, i.e. long range correlations. 
And similarly for links.  It does not fix he \system \ $\X$ uniquely. 
The requirement 1 says that emergent phenomena disappear when $\Phi_x$ 
are frozen. 

The quantities $\Phi_x$ were introduced into statistical mechanics by 
Kadanoff \cite{Kadanoff} under the name of {\em block spin}.
They are also  called {\em macros} \cite{macros}. The main 
problem in the approach is to find the suitable subsystems $X$ (blocks) and 
a good choice of block spins or macros.
 When one succeeds, the study of nonlocal 
phenomena - long range correlations - has been lifted to level 1, and 
a complexity reduction has been achieved. Now one can iterate the procedure. 

Given the blocks and a choice of block spin, how 
are the links in $\S^1$ and the cost function on level 1 constructed and
how can one find out whether the requirements 1 and 2 are satisfied? 

No general procedure is known which is always guaranteed to work. But 
the example below gives an idea how 
the task may be performed by enzymatic computation, at least 
in favorable cases. 

There remains the problem of how to choose the blocks and the block spins,
much as in the deterministic case. In successful applications
of the real space renormalization group to ferromagnets, lattice gauge 
theories
 \cite{Balaban}, and other problems in physics, successful
blocks and block spins could be guessed a priori. The big task for the 
future is to construct general and systematic procedures to find them.
Neural nets \cite{neuralNets} are a very difficult example of a prospective 
application. 

For a ferromagnet in thermal equilibrium, suitable blocks are 
cubes in space of some extension , 
and a suitable block spin is the total magnetization
in the cube. It fixes only the average value of the magnetic moment vectors
of the elementary magnets in the cube. But this is all that matters 
for the purpose of determining long range correlations i.e. the physics
at coarse scales \cite{Kadanoff}.

The construction of block spins is a cognitive procedure. It involves
construction of  {\em new} links which are not composed from existing
links alone, although very nearly so. In the construction below, the
irreducibly new link is the identification of $x$ with a representative 
(``typical'') object in $X$. 

In biological or social organisms we imagine that they have limited
cognitive capabilities which have been acquired by evolution. They determine
what blocks and blockspins are subject to being tried out. In the 
spirit of Ehresmann and Vanbremeersch \cite{Ehresmann}, one may 
imagine that they carry templates of the index category $\J$ which is
used in the construction of the block spin, cp. the example
 below and subsection
 \ref{sec:limit}.

\begin{example}{\em (Block spin)}
\label{example:blockspin}
The \system \ $\S $ and the cost functional $\H (\S )$ are the same as 
in the deterministic case, example \ref{example:multigrid}. 
$u_z \in \Omega_z$ are now random variables. Their a 
priori distribution is given by the uniform measure $du_z $ in $\Omega_z$ 
 and we seek to examine the probability measure
\ba
d\mu (u)&=& Z^{-1} e^{-\beta {\H}(u)}\prod_z du_z \\
{\H}(u) &=& \frac 12 \sum_{z,w} <u_z, L_{zw}u_w> + \sum_z <f_z, u_z>  
\ea
\end{example}
This would become (almost) a realistic Euclidean quantum field theory model of
strongly interacting elementary particles 
if $z$ formed a hypercubic 4-dimen\-sional lattice and 
if the lattice gauge fields $L_{zw}, z\neq w$ were
dynamical, with values in $SU(3)$. $d\mu $ is a Gaussian measure. 
\footnote{The qualification ``almost'' refers to the fact that 
$u$ should be Fermi fields rather than true random variables, and $L$ should
be the Dirac operator in a gauge field, which is not positive}    

Let us discuss the example. 
Consider subsystems $\X$ of $\bar{\S}$ (level 0), 
for instance a hypercube of some side 
length in the above mentioned 4-dimensional hypercubic lattice, together
with all links $L_{zw}$ between its objects (lattice sites) $z,w$. We seek to 
represent $\X$ by one object $x$ at the next level 1,
 and construct a block spin
$U_x \in \Omega_x$ as some average of $u_z$ over objects $ z\in X$,
\be U_x = \sum_{z\in X} C_{xz} u_z \ .\label{blockspinDef1}\ee
We omit $\circ$-symbols again for the composition of maps; $\sum $ is here 
addition in the vector space $\Omega_x$. 
$C_{xz}:\Omega_z \mapsto \Omega_x$ are linear maps. Elements $u_z\in \Omega_z$
for different $z$ 
are in different spaces. To add them up they need to be parallel transported
to some representative site $\hat x \in X$ first. This can be done with the 
help of a tree $\J$ with root $x$ which specifies a unique arrow $z\mapsto x$,
i.e. a linear map $t_{\hat{x} z}: \Omega_z\mapsto \Omega_{\hat x} $. $\J$
 could be a tree made of 
links of $X$ or it could be a star of arrows which are sums of 
parallel transporter from $z$ to $\hat x$ along some classes of paths. 
Given $\J$ we construct
\be C_{xz} = C_{x\hat{x}} t_{\hat{x} z} \ . \label{blockspinDef2}\ee
This involves one {\em new} link $C_{x\hat x}: \hat{x}\mapsto x$ which 
links the representative object $\hat{x}\in X$ in $X$ to the representative
$x$ of $X$ on the next level. Some such new link is inevitably needed since
we have no links between levels to start with. We choose $C_{\hat x x}$ as
identificaton map of $\Omega_{\hat{x}} $ and $\Omega_{x}$. This completes the
block spin definition.  

\begin{cat}
$\J $ is an unfrustrated subcategory of $Cat(\X)$ of the type of a 
partial order \cite{CWM}, p.11. At this stage we don't insist on making it 
into a \system , so adjoints of its arrows need not be in $\J$. But if 
the arrows $t_{xz}$ are unitary, we could put their adjoints into $\J$
and make it into a \system \ without introducing frustration. It 
retains the type of a preorder. 
\end{cat}

Now we come to the construction of the links at level 1 and 
the examination of the locality requirements. 

The standard procedure \cite{KG} is to construct an interpolation operator
 $A$ i.e. a collection of links $A_{zx}:x \mapsto z, \ z \in \bar{\S}$
for all $x\in \S^1$ in such a way that 
\be
(LA)_{zx} \equiv \sum_w L_{zw}A_{wx} = \sum_y  C^\ast_{yz} L^1_{yx} 
\label{LA=CastL1}
\ee
for some $L^1$. The links $A_{zx}$
 are supposed to be composed from $C_{xw}^\ast $ and 
arrows in $\bar{\S}$. The sum over objects $y$ of level 1 has only one term
 if every $z$ is in only one block $y$.

 A suitable  $A$ may be obtained by minimizing 
$\sum_{z,x} <A_{zx}^\ast, (LA)_{zx}> $ subject to the constraint $CA=1$, 
i.e. $\sum_z C_{xz}A_{zy} = \delta_{xy}$ for all $x,y$. This extra 
condition makes $A$ unique, and if the block spin choice is ``good'',
$A_{zx}$ will be local in the sense that $A_{zx}$ is very nearly zero 
except for $z$ in a reasonably small neighborhood of the subsystem $\X$. 
$L^1_{xy}$ are the desired links at the next level. Under the 
stated condition it will also be local - i.e. only a few
$L^1_{xy}$ will be not very nearly zero. This follows from 
$L^1 = CLA$ if $CC^\ast = 1$.

Given $A$, any $u$ can be uniquely split into 
contribution from a blockspin
$U=\{ U_x \}$ and a fluctuation field  $\zeta = \{ \zeta_z \}$ which satisfies 
the constraint $C\zeta = 0$ so that it contributes nothing to the block-spin.  
\be
u_z = \zeta_z + \sum_x A_{zx}U_x . 
\ee
The cost function decomposes as 
\be
{\H} (u) = \frac 12 <\zeta L \zeta > + <f, L\zeta > + \frac 12 <U,L^1U> + 
<Cf, L^1U> .
\ee 
The constraint $CA=1$ may be put into the measure $d\mu $ in the form 
of a $\delta$-function which is the limit of a Gaussian. One finds that 
$\zeta $ are Gaussian random variables \cite{GlimmJaffe} with covariance
$\Gamma = \lim_{\kappa\mapsto \infty} \Gamma^\kappa$, 
$\Gamma^\kappa = (L+\kappa C^\ast C)^{-1}$. One may opt to keep $\kappa$
finite, thereby relaxing $CA=1$. In this case $A=\kappa \Gamma_\kappa C^\ast $.

Then all locality requirements are fulfilled if $\Gamma^\kappa_{zw}$ decays 
fast with path-distance between $z$ and $w$. In particular, $A_{zx}$ and 
$L^1_{xy}$ will also be local.

In conclusion, one needs to find a local
``interpolation operator''  $A=\{ A_{zx} \}$
 such that 
eq.(\ref{LA=CastL1}) holds for some $L^1$. 
This yields  the links $L^1_{xy}$  of the 
\system \ $\S^1$ at the next scale.

\subsubsection{(Co)Limits}
\label{sec:limit}
Here I wish to establish the connection with the work 
in mathematical biology of 
Ehresmann and Vanbremeersh \cite{EV}.
 They propose to consider the objects $X$ in level
$j+1$ which represent subsystems of the level $j$ \system \ as {\em limits} 
in a category. The same construction of composite objects as limits is also
used in information science in what is called integration \cite{EhrigOrejas}. 
I will argue that (co)limits serve the same purpose as blockspins, and 
prove that the blockspin of section \ref{sec:RG} defines a colimit. 

I recall the notion of a limit \cite{CWM}.
 Given a category $\C$ and a (small) category 
 $\J$, called the indexing category, a functor 
$F:\J \mapsto \C$ is called a {\em diagram in $\C $ of type $\J$}.
 To be intuitive, 
Vanbremeersh and Ehresmann  call it a {\em pattern of linked objects}. 
By the map, some objects $F_j=F(j)$ of $\C $ are indexed by objects $j$ of 
$\J$. 

An object $L$ of $\C$ together  a collection of arrows $\pi_j: L \mapsto F_j$,
one for each $j\in J$, is called a {\em cone} $\pi: L\mapsto F$
on the diagram $F$ of type $\J$ with vertex $L$
 if the following 
compatibility condition is satisfied. For any arrow $u:i\mapsto j$ in $\J$, 
\be \pi_k = F(u)\circ \pi_j . \label{collectiveLink} \ee
In ref. \cite{EV}, $\pi = \{ \pi_j \}$ is called a {\em collective link}. 
It links $L$ to the subsystem ${\bf L}$ which it is 
supposed to represent. In our applications, the subsystem has $F_j$ as 
its objects, but in general it has more arrows than $F(u), u\in \J $. The 
images of 
links $u \in \J$ are special links, which are
 ``important for the collaboration``. Note that the collective link 
projects out any frustration in the image of $\J$ in the following sense.
Given two arrows $F(u_1):F_i\mapsto F_j$ and  $F(u_2):F_i\mapsto F_j$, 
eq.(\ref{collectiveLink}) implies 
$  F(u_1)\circ \pi_j = F(u_2)\circ \pi_j \ . $.
If $F$ maps several $j$ to the same object $z=F(j)$, we need only one 
link $\pi_j \equiv \tilde \pi_z: L\mapsto z$ for them, 
and we may regard $\tilde \pi_z$ as 
a (link-) field whose argument is $z\in {\bf L}$. The condition 
(\ref{collectiveLink}) can be interpreted to say that 
this field is constant under parallel transport 
along paths which are images of paths in $\J$. We say that it is ``constant
 along $\J$'' for short.  

A cone $\pi : L \mapsto F$ 
is called a {\em limit of the diagram $\J$}
 if the following uniqueness property holds. 
Given any other cone $f: Y \mapsto F$, there exists a {\em unique}
arrow $g:X\mapsto L$ such that $f_j = \pi_j \circ g $. 
By abuse of language, $L$ is called the limit. 
 The condition of a cone $f$ means that $\tilde f_z$ is constant 
along $\J$.

Colimits are dual to limits, i.e. all arrows are reversed.  
In our applications, all categories except the index categories $\J$
will be \system s. Therefore $\{ \pi_j \} $ defines a limit if 
$\{ \pi^\ast \}$ defines a colimit
of a dual diagram of type $\J^{op}$, 
where $\J^{op}$ is $\J$ with arrows reversed \cite{CWM}. 
\begin{theorem}
\label{theo:colimit}
{\em (Block-spins as colimits)}
The block-spin construction of section \ref{sec:RG} based on a tree $\J$ 
defines  a colimit $x\in \S^1$ in the category which
contains $Cat(\S)$, $Cat(\S^1)$
and the links $C_{x,z}: z \mapsto x \ , (z\in \S)$, and in any category
containing it.
\end{theorem}
The interpretation of block-spins as (co)limits serves to translate from a
quantitative description to a structural one. 
Before we proceed to the easy proof of theorem \ref{theo:colimit},
 let us discuss how the interpolation 
operator is interpreted. Equation (\ref{LA=CastL1}) requires that for every $x$,
\be \tilde f_z = (LA)_{zx} = \sum_w L_{zw}A_{wx}: x \mapsto z 
\label{factorCone}
\ee
is constant along $\J$, because this is true of $\tilde \pi_z = C^\ast_{zy}$. 
In other words, 
$\{ (LA)_{zx} \}$ is a collective link which defines a cone on the same
diagram as for the collective link $\{ C^\ast_{zx} \}$. 

If every object $z$ is in only one block $\hat z$, then 
$(C^\ast L^1)_{zx} = C^\ast_{z\hat{z}}L^1_{\hat{z}x} \mbox{ (no sum)}$.
 Therefore,
if $LA $ defines a cone, the existence of $L^1$ is assured by the property 
that the collective link $\{ C^\ast_{zx} \}$ defines a limit. 

In conclusion, the existence of the interpolation operator requires the 
existence of another ``factorizing'' cone on the same diagram as the limit 
cone. The factorization into $L$ and some $A$ is expressed by 
eq.(\ref{factorCone}), $L=\{L_{zw}\}$ is the collection of links of the 
\system  \ $\S$ at the fine scale. 
(Note that this is not of the form of standard
factorization properties in category theory because of the sum. It involves
the $\oplus$-operation of semiadditive systems.) In addition, the locality 
properties laid down in previous subsections should be satisfied. This means
that it must be possible to approximate $A_{zx}$ by zero if $z$ is not in a 
reasonably small neighborhood of the subsystem $\X$ represented by $x$. 
How to go about treating the error made in this way is a subtle issue 
\cite{gupta:trunc} which 
I am not prepared to discuss here.

Let us turn to the proof of the theorem. 
The definition of a limit is external in the sense that one needs to seek an
arrow
 in the whole category and show its uniqueness. This is typical of  the 
universal constructions of category theory. This feature is what makes
category theory into ``abstract nonsense''. But there are instances where
limits are internal, i.e. require only examination of the arrows which are
involved in their construction. Products in {\em Ab}-categories are examples.
They are necessarily biproducts, and biproducts are easy to characterize
internally, cp. theorem 2 of section VIII in \cite{CWM}. 

Products are limits with index categories $\J$ which have no arrows other than 
the identity arrows. A similar situation holds when $\J$ is a tree or
 preorder. 
\begin{lemma}
\label{lemma:tree}
{\em (Colimits on trees)}
Let $\J$ be a tree with root $r$, so that it is unfrustrated and there
is a unique arrow $t_j:j\mapsto r$ in $\J$ for every object $j\in \J$. Then 
a cocone $\pi: A(\J) \mapsto L$ is a colimit if $\pi_r$ has an inverse.
Conversely, the colimit property requires that 
$\pi_r$ has a left inverse $i_r$, viz $i_r \circ \pi_r = \iota_L $.  
\end{lemma} 
The dual statement is true for limits. 

{\sc proof} of theorem \ref{theo:colimit}.
$\J$ and $F(\J)$ are identified in section \ref{sec:RG}. 
The collective links are $C_{xz}: z\mapsto x$, $(z \in A(\J)),$
 and the root is 
$\hat x$. The compatibility condition for a cone is satisfied by
 construction.
 The theorem is an immediate
consequence of the lemma, since $C_{x\hat{x}}$, which substitutes for $\pi_r$, 
was chosen as an identification map, whose adjoint is its inverse. q.e.d.

{\sc Proof} of lemma \ref{lemma:tree}. 
Given another cocone $f$, $g=f_r \circ i_r $ is the required unique map. 
It is unique because $f_r = g\circ \pi_r = h\circ \pi_r$ implies $g=h$ by
invertibility. 
Conversely, $t={t_j}$ is a cocone with vertex $r$ and the required unique
map $g$ must be a left inverse of $\pi_r$. 
\begin{define}{\em [Ehresmann Vanbremeersch]} 
A {\em hierarchical system} is a category $\HC$ whose objects are divided
into {\em levels}, numbered $0,1,...,p$, such that each object of level $n+1$
(where $n<p$) be the limit in $\HC$ of a pattern $A$ of linked objects
$[$=diagram$]$ of level $n$ (i.e. each $A_i$ has level $n$). 
\end{define}
I propose to substitute ''\system `` $\HC$ for ``category $\HC$'', and 
count the arrows $\pi_i $ in the collective links as links.  I would also
prefer to speak of colimits in place of limits. 

If indeed blockspins and colimits are basically the same, as is suggested 
by the above theorem \ref{theo:colimit}, this setup corresponds with the
 multiscale analysis of section \ref{sec:multiscale}. 

It is always possible to extend categories by adding objects which represent 
limits of certain diagrams \cite{Ehresmann}. 
Mathematicians often speak of categories which 
have all finite limits (i.e. limits for all finite diagrams). But to do so
would be contrary to the intended complexity reduction. 

\subsection{Dynamics on coarser levels}
\label{sec:coarseDynamics}
The somewhat abstract considerations of  section \ref{sec:limit} 
serve to convert 
blockspin constructions from the quantitative description that is 
used in quantum field theory and statistical mechanics to a structural
 description.  

The problem is now how to extend the dynamics from level $0$ to the higher 
levels. We think of a stochastic dynamics. In autopoietic systems, 
the objects in the higher levels will typically  represent functional units 
of objects of
lower levels which should be capable of making their elements.
 They may disappear, i.e. die. They may live on, possibly 
adapting or differentiating. And new ones may form, 
for instance as newly made copies of already existing units, building blocks 
being absorbed from the environment.  
How and when does this happen?

Diagrams of type $\J$ were identified with certain types of block spins,
 and the object $x$ to which the block spin is attached 
was identified with a limit of the 
diagram. 
The objects $x$ represent  subsystems $\X$ of $\S$ whose constituents 
cooperate through links (channels of communication) which are 
determined by the 
diagram. $x$ may stand for 
organs in a biological organism, for institutions of a society, for extended
domains in a ferromagnet, or for any kind of a ``thing''.
 We only want to  keep or acquire them 
in our model when they achieve
 something or are needed to achieve something, namely the cure of locality 
problems as discussed in sections \ref{sec:MGrid},\ref{sec:RG}. 
This is the criterium by which objects representing subsystems will appear or 
disappear.

An object $x$ of this kind in level $n+1$ may disappear if the
 limit of a diagram in level $n$ ceases to exist.
This mechanism was suggesed by Ehresmann and Vanbremeersh. It can happen when
the diagram is disrupted because the links involved in the collaboration 
disappear. I give an example in a moment. It may also cease to exist because
frustration appears in the diagram, so that the cone ceases to exist. 
Informally speaking, confusion arises because communication of the 
collaborators through different channels produces different messages.
(If ${\J}$ is a tree, this cannot happen).

It may occur that there are short range correlations only 
on level $n$ in the vicinity of some subsystem $\X$, even 
without any blockspin constraint. No block $x$ 
on level $n+1$ is needed in
that case. If it is present anyway, it will have no important links 
$L^{n+1}_{xy}$ 
to other objects $y$ on level $n+1$, hence no relations 
to anything in the ``rest of the world $\S^{n+1}$''.
 Such an object does not exist 
for the ``rest of the world'' and should be discarded.

Conversely, if locality properties in $\S$ or in time start to be violated, 
they need to be salvaged by a introducing a blockspin constraint as discussed 
in sections \ref{sec:MGrid},\ref{sec:RG},
 and with the block spin comes an object 
to which it is attached. 
Suppose that the correlations are not short ranged at 
level $n$ without block spin constraint.
This can be decided on the basis of relaxation sweeps,
i.e. by enzymatic computation. They  
will produce correlations growing beyond the allowed range. In this case,
suitable blocks and block spin constraints will have to be introduced 
until the residual correlations under the block spin constraint are
short ranged. If the multiscale analysis were done on the drama,
short ranged would mean in particular short ranged in time. Thus, the 
effect of particular properties (initial conditions) 
which are independent of the value of the  blockspins will die out quickly. 
Only the cooperative effects which are well described by the block spins
will survive.

Here is the example for the loss of a diagram. 
 It is a 
frequent cause of death that an organism or a functional part of it
gets digested by a predator or parasite. Digestion is performed by special 
enzymes. For instance, T4 bacteriophage's nuclease enzymes degrade its 
{\em E.coli} host's chromosome (but not its own genome).  
 Consider one organism ${\bf P}$ being digested by another one
who attacks it by acting on its object $X$ with its digestion enzyme
according to the model mechanism of section \ref{sec:digestion}.
Take any subsystem ${\bf Q}$ of ${\bf P}$ which does not contain $X$.
It will be totally disconnected after the digestion process. Therefore,
the supposed images of links in the diagram $\J$ - the links which  are essential for the collaboration - are missing.  
 So the loss of structure at some level,
e.g. by digestion, may lead to loss of the limit object in the next level.

The choice of block spin will typically not be unique. But 
if  suitable block spins can be found, the long range correlations are 
under control, and therefore all emergent phenomena. They are merely
described in a different language when different block spins are chosen. 

What has been proposed here is a reductionist scenario - the dynamics
on the lower level determines what happens at higher levels, modulo
switch of language. Locality is the crucial ingredient of the construction. 
Earlier multi-level analyses of biological systems \cite{EV,Baas:1,Emmeche}
 had no locality principle and they relaxed on reductionism.
Extreme views were expressed by Laughlin and Pines. They claimed that neither life nor high temperature superconductivity can be understood from basic principles \cite{LaughlinPines}.  

Autopoietic systems make their own elements. This appears to require a 
top down action of objects $x$ at level $n+1$
 to make elements which are objects
or links at level $n$ inside the subsystem $\X$ to which $x$ corresponds.
We understand this now. There is a dynamics at level $n$  which gives rise 
to nonlocal - therefore emergent - phenomena.
 And these are effectively described with the help of objects $x$ at
 level $n+1$. 

I add few words on funtionality. 
Technically most convenient would be a multiscale analysis of the drama.
In this way, the higher levels would also acquire longer time scales,
and function would appear as structure. 
For intuition's sake, I spoke here of changes in time instead. Therefore we 
needed to speak of functionality separately. 
In the present approach, functionality depends on the presence of a suitable
 complement of enzymes. The integrity of boundaries of subsystem may also be 
important in order to confine the domain where enzymes act, and also the presence of 
channels of communication which transfer quantities of material constituents.
All this should be reflected at the coarser scale when the functionality is 
important for cooperation at that scale. 

Production of copies is an emergent phenomenon, 
as we saw in section \ref{sec:splitFork}. Typically the copy process
absorbs objects which involve, at a still lower level, materials 
(including energy)
which are conserved or supplied by the environment in limited 
quantities. In this way, a competition for scarce resources results which 
drives evolution.

Imagine a subsystem $\X$ is copied by the splitFork dynamics because 
a sufficient collection of  microenzymes is present in $\X$. 
Then the splitFork-enzyme,
now considered as one entity, should be attached to the object $x$ 
which corresponds to $\X$ at the next scale.   

Let me emphasize that a general block spin procedure can be much more 
complicated than for a ferromagnet, where the Kadanoff construction 
furnishes one block spin definition for all purposes and all scales. In 
general, the appropriate {\em kind} of blockspin on level $n+1$ may depend on 
the {\em values}
of variables or block spins $\varphi$ on level $n$. Since $\varphi $ are 
random variables, there may be nonvanishing probabilities for 
the appropriateness of several kinds of block spin descriptions. This 
can lead to bifurcations. The situation on the next scale may be the same 
again, and a proliferation of possibilities may result. One cannot expect
to find giraffes by enumeration of all possibilities of living organisms.
Moreover, scaling laws (power laws) are expected to emerge  in special
 circumstances only. Ferromagnets, the Bak Sneppen model of evolution
\cite{BakSneppen} and Lotka-Volterra-models \cite{LotkaVolterra}
are examples, but power laws are not a general indicator of criticality.  

\subsection{Deductive vs. observational emergence}
Baas \cite{Baas:1} makes a distinction between deducible and observational
 emergence. He interpretes G\"odels incompleteness theorem in logic
as a case of observational emergence.

In the present frameork, we may classify as {\em deduction}
in this sense anything that 
involves composition of existing links (with $\circ$ and $\oplus$).
This does not change the category. Deductions in proposition logic are of
this kind, cp. section \ref{sec:logic}, figure \ref{fig:logic}.  
Observational emergence would then involve the making of cognitive links.
They are new links added to the category. 
The blockspin constructions involve new links. 
\section{Semicommutative differential calculus and geometry on \system s} 
\label{sec:NCDG}
In this section I want to bring \system 's theory closer to the traditional 
approach in physics which is based on differential calculus.

Given a \system \ $\S$, there is a unique local functor to  an unfrustrated 
\system \
$\B$ which shares the objects with $\S$ and inherits equivalence classes 
of its parallel links.  In $\B$, different links of $\S$ with 
the same source and target are identified, and equally the arrows. This 
is an example of the possible identifications mentioned in section 
\ref{sec:tauto}.  $\B$ is determined by the graph of $\S$ and 
shall be called the {\em base} \system\ of $\S$,
or {\em base} for short. I assume for simplicity that it has at most countably 
many elements 
\footnote{$\B$ need not be like a grid. For instance, it might contain 
grid-like subsystems together with arbitrary refinements of these. The 
floating lattices in the Ashtekar approach to gravity are like this
\cite{Ashtekar}}.

Differential calculus and geometry puts more structure on a given base, 
thereby creating \system s $\S$ with {\em given} base, 
and it constructs algebras and modules to go with them. 
Changes in connectivity of the base are outside the scope of calculus.

The appropriate version of differential calculus to fit onto arbitrary
unfrustrated \system s $\B$ is a special case of noncommutative 
differential calculus and geometry \cite{connes} which was 
developped by Dimakis and M\"uller-Hoissen \cite{dimakis}. I call it 
semicommutative because the ``algebra of functions''
${\cal A}$ remains commutative. In this framework, conventional notions
of locality and the notion of a point, which are given up in fully 
noncommutative differential calculus, retain their meaning. 
One usage is in lattice gauge theory on a hypercubic lattice. 
All the familiar formulae from gauge theory in the continuum remain 
literally true, except for the commutation relations of differentials 
with functions \cite{dimakisLatt},  cp. eq.(\ref{fdx}). 

The use of this device is in the spirit of the strategy to bring proven 
methods of theoretical physics to bear on very general complex systems.
The discrete calculus substitutes for and is in many ways like calculus on 
manifolds. 

Given a base $\B$, let ${\cal A}$ be the algebra with unit element
consisting of real or complex functions on $\B$ with pointwise
multiplicaton. 
It has a basis $\{ e_X\}$   labeled by objects $X$ of $\B$ such that
\be e_X e_Y = \delta_{XY} e_X \ee
$\delta_{XY}$ being the Kronecker $\delta$ -function

From now on, the symbol $X\mapsto Y$ shall mean that there exists a
 link from $X$ to $Y$. Since there is 
 at most one such link $b$
we may write $b=(XY)$. The differential algebra $\Omega $ is generated by 
the algebra ${\cal A}$ of functions and differentials $e^{XY}$ attached to 
the links of $\B$. $\Omega_1=span_{\bf C}\{ e^{XY}: X\mapsto Y \}$
is made into a ${ \cal A}$-bialgebra via
\be
e^Ze^{XY} = \delta^{ZX} e^{XY}, \  e^{XY}e^Z = \delta^{YZ} e^{XY}. \ 
\ee
The quantity $\rho = \sum e^{XY} $ (sum over all links)  is introduced,
and exterior differentiation $d$ is defined by 
\ba de^X &=& \rho e^X - e^X\rho , \\
de^{XY} &=& \rho e^X \rho e^Y - e^X \rho^2 e^Y + e^X \rho e^Y \rho 
\ea
and the standard graded Leibniz rule. 

There are relations between the generators in the algebra. 
Whenever the link from $X$ to $Y$ is missing,  $e^X\rho e^Y=0$.
Applying $d$ implies the constraint $e^X \rho^2 e^Y=0$, i.e. 
\be
\sum_Z e^{XZ}e^{ZY}=0 \ .
\ee  

It was shown by Dimakis and M\"uller Hoissen that this calculus reduces to 
something looking familiar on an ``oriented'' d-dimensional hypercubic lattice
of lattice spacing $a$. 
Of the two directions $\pm \mu $, one is distinguished as positive,
 say $+\mu$,
$\mu = 1,...,d$, and links are
put between nearest neighbors denoted $x $ and $x+\mu $  in positive
 direction only. 
 If $x^\mu $ are the standard coordinate functions, 
one computes
 $dx^\mu = e^{x,x+\mu}$ and therefore $df(x) = a^{-1}[f(x+\mu)-f(x)]dx^\mu $.

The constraints among differentials reproduce 
the standard relations $dx^\mu dx^\nu + dx^\nu dx^\mu = 0 $
There is a
Hodge * -operator, and the calculus shares all the properties of the 
continuum calculus, except that 
\be 
 f(x)ax^\mu = dx^\mu f(x+\mu )  \label{fdx}
\ee 
This rectifies the Leibniz rule. To understand  the usefulness of all this,
consider

\begin{theorem}{\em (Gauss constraint)}
The Maxwell dynamics of a free electromagnetic field in discrete space and
time preserves the Gauss constraint. 
\end{theorem}  
{\sc Proof:} One transcribes the standard proof. 
Let $d_s$ be the exterior derivative in space, $*$ the Hodge star operator 
in space and $d_s^\ast = * d * $.
The electric field defines a 1-form $E=\sum E_i dx^i $.  
The Gauss law says that $d_s^\ast E = 0.$   
The magnetic field  defines a 2-form $B$ and the equations of motion says that
$\dot E = d_s^\ast B $. Since $ d^{\ast 2} = 0$ it follows that 
$d^\ast_s \dot E = 0$, and so the Gauss constraint is preserved. All
the quantities have their analog in the semi-commutative calculus on an 
oriented cubic lattice, $\dot E$ becomes the finite difference derivative,
and the equations of motion and Gauss constraint retain their form. So
the proof carries over.  

\section*{Acknowledgement}
I would like to thank A. Brandt, I. Cohen, M. Meier-Schellersheim, S. Solomon,
J. Wuerthner and Y. Xylander for many helpful and stimulating discussions.

\end{document}

%% file: catalyse.pstex_t
\begin{picture}(0,0)%
\epsfig{file=catalyse.pstex}%
\end{picture}%
\setlength{\unitlength}{4144sp}%
\begingroup\makeatletter\ifx\SetFigFont\undefined%
\gdef\SetFigFont#1#2#3#4#5{%
  \reset@font\fontsize{#1}{#2pt}%
  \fontfamily{#3}\fontseries{#4}\fontshape{#5}%
  \selectfont}%
\fi\endgroup%
\begin{picture}(5750,968)(316,-635)
\end{picture}